\documentclass[twocolumn,english,aps,prb,showpacs]{revtex4-1}

\usepackage{graphicx}

\usepackage [latin1]{inputenc}

\usepackage{amsmath}
\usepackage{color}
\usepackage{epstopdf}

\begin{document}

\title{Inhomogeneous micro-strain in cylindrical semiconductor heterostructures and its influence on the adiabatic motion of electrons}
\author{B. Jenichen}
\email{bernd.jenichen@pdi-berlin.de}
\author{U.~Jahn}
\author{A.~Nikulin}
\email{on leave from Monash University, Clayton, Victoria 3800, Australia}
\author{R.~Hey}
\author{P.~V.~Santos}
\author{K.-J.~Friedland}
\affiliation{Paul-Drude-Institut f\"ur Festk\"orperelektronik,
Hausvogteiplatz 5--7, 10117 Berlin, Germany}

\date{\today}

\begin{abstract}
We analyze fluctuation of the layer thicknesses and its influence on the strain state of (In,Ga)As/(Al,Ga)As micro-tubes containing quantum well structures. In those structures a curved high-mobility two-dimensional electron gas (HM2DEG) is established. The layer thickness fluctuation studied by atomic force microscopy, x-ray scattering, and spatially resolved cathodoluminescence spectroscopy occurs on two different lateral length scales. On the shorter length scale of about 0.01~$\mu$m, we found from atomic force micrographs and the broadening of the satellite maxima in x-ray diffraction curves a very small value of the mean square roughness of 0.1~nm. However, on a longer length scale of about 1.0~$\mu$m, step bunching during epitaxial growth resulted in layer thickness inhomogeneities of up to 2~nm.
The resulting fluctuation of the strain in the micro-tubes leads to a local variation of the chemical potential, which results in the fluctuation of the carrier density as well. This leads to a phase cancelation of the Shubnikov-de-Haas oscillations in the curved HM2DEG and a  reduction of the single-electron scattering time, while the electron mobility in the structures remains high. The estimated fluctuation of the carrier density agrees well with the energy fluctuation measured in the cathodoluminescence spectra of the free-electron transition of the quantum well.
\end{abstract}


\maketitle

\section{Introduction}

\begin{figure}[!t]
\includegraphics[width=8.0cm]{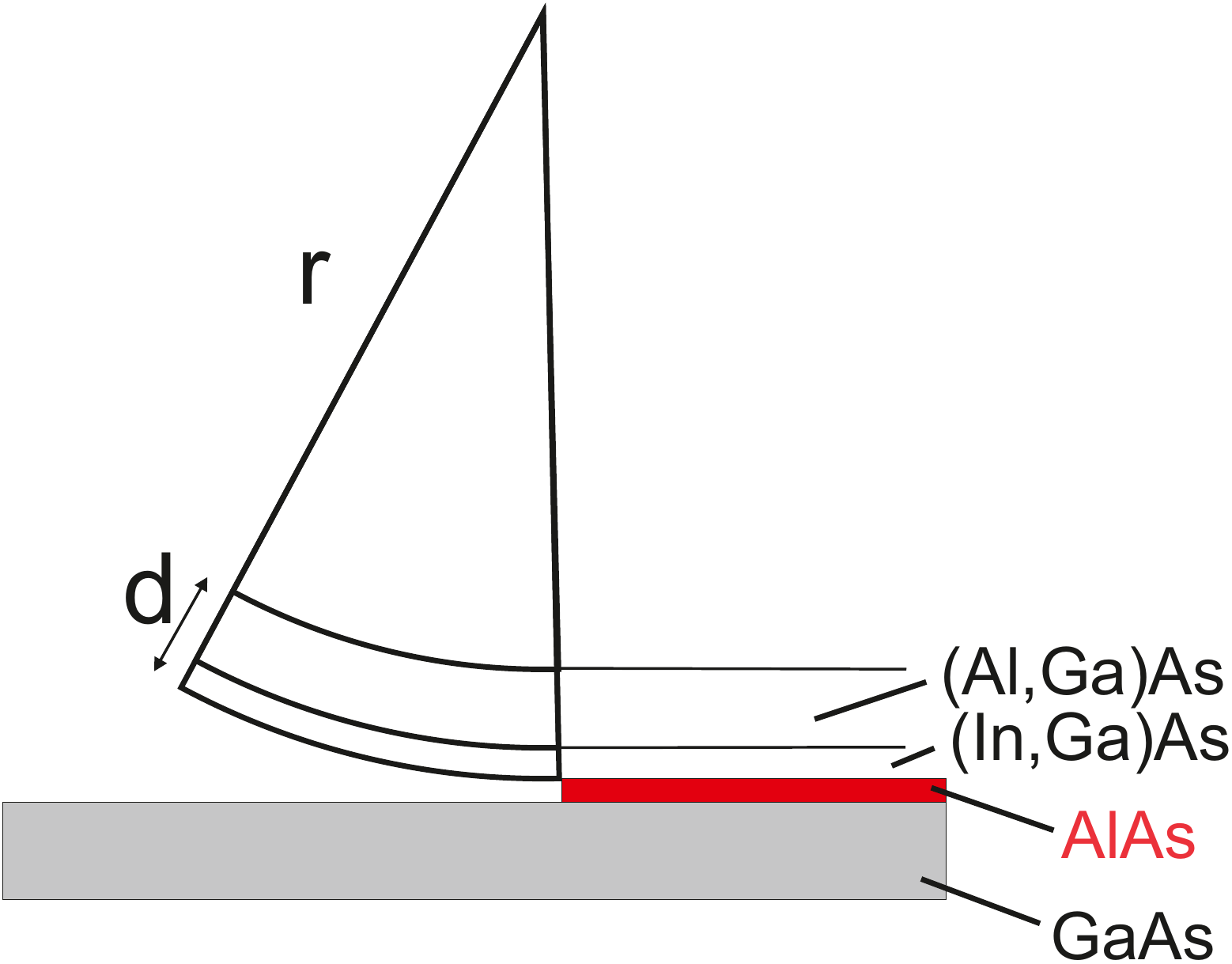}
\caption{ (Color online) Schematic view illustrating the self-rolling of a layer stack consisting of an (Al,Ga)As/GaAs QW structure and an (In,Ga)As stressor film. The barriers consist of AlAs/GaAs SLs. After etching away the sacrificial AlAs film between the layer stack and the thick GaAs substrate, the elastic stress relaxation by bending is not influenced any more by  the substrate.}
\label{fig:bimetall}
\end{figure}

The self-rolling of thin pseudomorphically strained semiconductor bilayer systems
based on epitaxial heterojunctions grown by molecular-beam epitaxy (MBE)
was proposed by Prinz et al.\cite{prinz00} Such structures allow for investigation of the physical properties
of systems with nontrivial topology. A new research field was opened: It became possible to create a high-mobility two-dimensional electron gas (HM2DEG) on cylindrical surfaces. In these structures, the low-temperature mean free path of the electrons $l_{\text{MFP}}$ can be kept comparable to the
curvature radius $r$.\cite{friedland2007, vorobev2007} In this case, the modifications of the adiabatic
motion of electrons on  cylindrical surfaces lead to trochoid- or snake-like trajectories~\cite{friedland2007} or additional structures in the quantum Hall effect.\cite{Friedland2009}
The local structure of rolled-up  single crystalline structures was already investigated in detail by x-ray micro-diffraction.\cite{Krause2006,Malachias2009}
A review about the structure of radial superlattices is given in Ref.(\onlinecite{Deneke2009}).
Here we demonstrate that lateral strain fluctuation arising during the stress relaxation by the release from the substrate and the self-rolling of the  strained semiconductor bilayer systems significantly affects the adiabatic motion of the electrons on curved surfaces.
Figure~\ref{fig:bimetall} illustrates the self-rolling of an (Al,Ga)As/GaAs quantum well (QW) structure and an (In,Ga)As stressor film after etching away a sacrificial AlAs film underneath the (In,Ga)As film.
The originally tetragonally strained (In,Ga)As film can now expand laterally (parallel to the hetero-epitaxial
interface) by curving the whole remaining stack, i.e. the thin (Al,Ga)As/GaAs  QW structure grown epitaxially
on top of the stressor film, like a bimetal strip. $d$ is the thickness of the
layer stack. The radius $r$ of the cylindrical structure is determined by the thicknesses
and elastic moduli of its individual layers.\cite{Grundmann2003}  Simultaneously, the
self-rolling process itself creates a new strain topology in the layers. First, a gradient
exists ranging from compressive to tensile strain in a rolled-up layer stack  normal to the
layer surface. The difference between the strain at the upper and lower surfaces of the stack  is
$\Delta_\bot\varepsilon= d/r$.
Furthermore, a lateral fluctuation of the local strain $\Delta_L\varepsilon$ is caused by
tiny  variations of the layer stack thicknesses $\delta d$, which
arise on different length scales. On a lateral length scale of several micrometers, thickness fluctuation occurs due
to step bunching during the MBE growth procedure with  peak-to-peak height differences of 1--2 nm.\cite{Kondrashkina1997,jenichen1997}
Consequently, for  a low value of $d$, this leads to an essentially inhomogeneous strain field penetrating the structures. The thickness fluctuation has a large influence on the electronic properties of self-rolling systems, which lack the stabilizing influence of a thick substrate.
This thickness fluctuation and the corresponding strain inhomogeneities cause
lateral changes of the energy~gap of the structures leading to variations of the quantum
well confinement energies.\cite{jahn1994,jahn1995,jahn1996}
Consequently, for the respective cylindrical structures with a radius of $r~=~20~\mu$m, we should expect   an overall strain fluctuation of
 $\Delta\varepsilon~=~\delta d/r~=~1\times10^{-4}$, which translates
 into  energy fluctuation up to $\Delta E=|a_{\text{CB}}|\cdot\Delta\varepsilon~\simeq~1$~meV (deformation potential $|a_{\text{CB}}|~\simeq~10$~eV~\cite{yu1995}).
In the present work, we show  that layer thickness fluctuation on a nanometer scale
leads to a phase cancelation of the Shubnikov-de-Haas (SdH) oscillations in the curved HM2DEG.

\section{Experiment}

 Strained, multilayered films (SMLF) with overall thicknesses of $d_{L}~\cong$~150--170~nm were grown by MBE on top of a nominally $d_S$=24~nm  thick
(In,Ga)As stressor layer.   Each of the SMLF included a HM2DEG in a remotely doped GaAs quantum well (QW) structure. The barriers of the QW consisted of AlAs/GaAs short-period superlattices (SL) with Si-$\delta$ doping in  one of the GaAs sequences in the SLs on both sides of the QW. This specific structure allows for a high electron mobility  even in  freestanding layers.\cite{friedland1996,friedland2001}
The surface morphology of the flat samples was studied by atomic force microscopy (AFM)
in ambient air using a Digital Instrument Nanoscope.
High-resolution x-ray diffraction (XRD) measurements were performed on the flat as-grown structures
using a Panalytical X-Pert PRO MRD\texttrademark\ system
with a Ge(220) hybrid monochromator  (Cu~K$\alpha_1$ radiation with a
wavelength of $\lambda=1.54056$~\AA). The program
Epitaxy\texttrademark\ was used for data evaluation. $\omega$ is the angle between the incident x-ray beam and the sample surface. 2$\Theta$ is the detector angle with respect to the incident beam.

The depth profile of the displacement of the original layer stack was calculated directly by the phase retrieval method \cite{nikulin1998,nikulin1999} from synchrotron x-ray diffraction data. The synchrotron experiment was conducted on BL13XU at SPring-8. An  x-ray energy of 12.4 keV was selected by the primary Si(111) channel-cut beamline monochromator. Further collimation and monochromatization of the beam were performed using a secondary channel-cut Si(400) monochromator. We recorded diffracted intensity profiles as a function of the sample angle and location of the incident beam on the sample surface. The experimental parameters allowed us to achieve a spatial resolution in the resulting strain depth profiles of 0.5~nm. Plane wave illumination was performed using a 20~$\mu$m wide slit in the direction of the diameter of the roll. The other dimension of the beam (perpendicular to the diffraction plane) was 100~$\mu$m. The location of the incident beam on the sample surface was changed with a linear step of 5~$\mu$m.

For the investigation of curved HM2DEG's, we first fabricated conventional
Hall bar structures in the planar heterojunction  along the $[100]$
crystal direction with  three
4~$\mu$m wide lead pairs, separated by 10~$\mu$m  in a
similar manner as used in Ref.~\onlinecite{Friedland2009}. Then, the SMLF
including the Hall bar was released by selective etching of the
sacrificial AlAs layer with a 5$\%$ HF acid/water solution at 4~$^\circ$C starting from a $[010]$ edge. In order to relax the
strain, the SMLF rolled up along the $[100]$ direction forming a
complete tube with a radius $r$ of about 20~$\mu$m.  The direction of the current in the Hall bar was along the circumference of the tube. For our measurements of the SdH oscillations in the longitudinal resistances, we thoroughly aligned the direction of the magnetic field perpendicular to the cylinder surface at the position in the middle between the terminals for the resistivity
measurements by minimizing the asymmetry effects in the magnetotransport data,
 caused by the  magnetic field gradient  $\nabla \textbf{B}$.\cite{Friedland2009} All resistance measurements were  carried out at  temperature  $T$~=~50~mK. We determined a mobility $\mu~=~100~$m$^2~$V$^{-1}~$s$^{-1}$ and a carrier density of $n~=~5\times10^{15}$~m$^{-2}$.

Spatially resolved measurements of the luminescence of the cylindrical structures
were performed by cathodoluminescence spectroscopy (CL) in a scanning electron microscope (SEM).
Secondary electron (SE) and monochromatic CL images were acquired simultaneously for an accurate
assignment of the local origin of the CL. The CL/SE experiments were performed using a Zeiss ULTRA55
field emission SEM equipped with the Gatan monoCL3 and a He-cooling stage system. The acceleration voltage was 5~kV while the beam current was 5~nA at
a sample temperature of 7~K. The spectral resolution amounted to 0.3~nm corresponding to a slit width of the CL spectrometer of 0.1~mm.

We investigated always samples from the same MBE growth run by all the different methods. In our experience the inhomogeneity of the MBE grown wafer in terms of composition and layer thickness is below 5\%. An error of 5\% can be tolerated for the considerations in the present work. From the MBE grown samples many rolls were prepared and some of them were investigated by SEM, others by transport measurements. We assume, that the roughness of the interfaces arises during the MBE growth procedure and is not changed by the rolling. From  XRD and  AFM we obtain quantitative data of the root mean square (RMS) roughness according to standard procedures. In addition we obtain from AFM quantitative data of the peak-to-peak height values in typical AFM micrographs. All this data is obtained from as-grown samples.  The AFM and x-ray data are representative for the step flow growth mode of MBE [see Ref.(\onlinecite{Kondrashkina1997,jenichen1997}) and references therein].

\section{Results and Discussion}
\subsection {Surface and interface roughness}
\begin{figure}[!t]
\includegraphics[width=8.0cm]{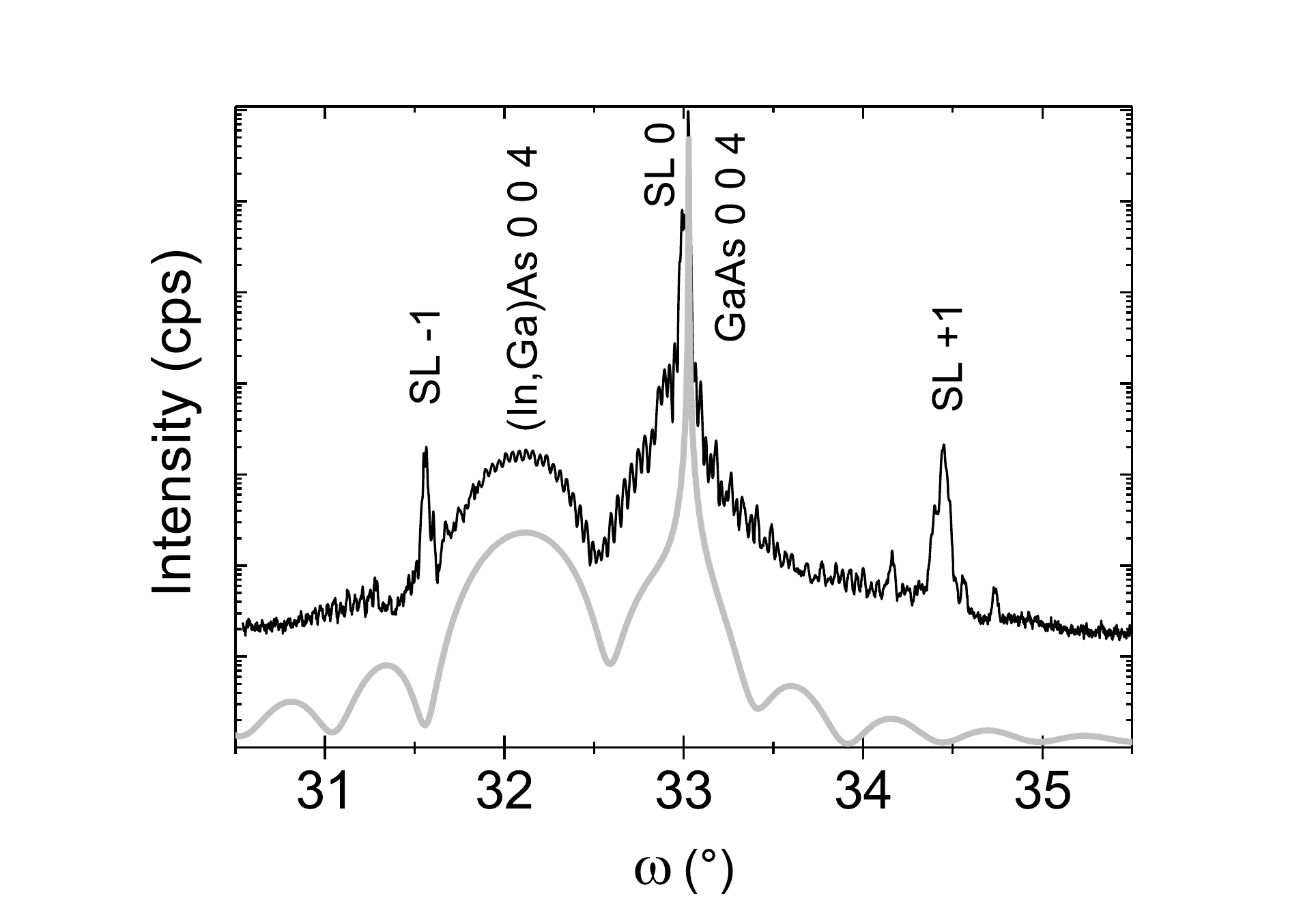}
\caption{ XRD curve of the as-grown (Al,Ga)As/GaAs QW structure with AlAs/GaAs SL barriers and an (In,Ga)As stressor film near the 004 reflection. The satellite reflections of the short-period SLs are marked as well as the (In,Ga)As peak and the GaAs substrate reflection. The simulation on the bottom was performed for a 10~nm thick In$_{0.19}$Ga$_{0.81}$As stressor layer on GaAs in order to illustrate the main source of strain in the structure}
\label{fig:laboratory}
\end{figure}

\begin{figure}[!t]
\includegraphics[width=8.0cm]{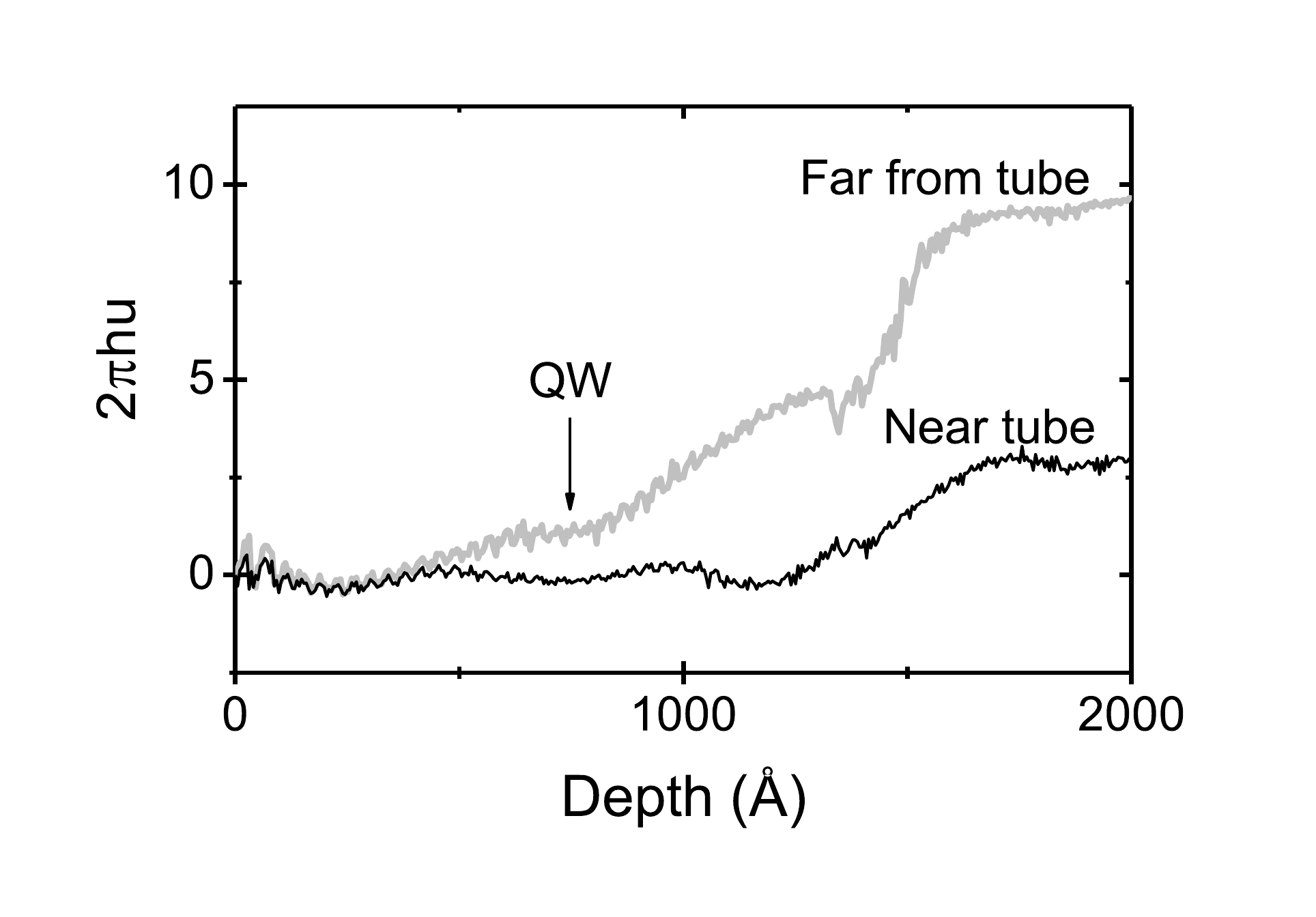}
\caption{   Displacement profiles obtained by the x-ray  phase retrieval method far away from a micro-tube and in the vicinity of a micro-tube. The corresponding measurements (not shown here) were performed with an intense, highly collimated, low-diameter synchrotron beam. $u$ denotes the displacement, and $h$ is the magnitude of the reciprocal lattice vector of the corresponding reflection (004). The position of the GaAs QW is marked.}
\label{fig:PR}
\end{figure}

The RMS thickness fluctuation in multilayer structures can be detected by XRD via a broadening of their SL
maxima.\cite{speriosu1984,holy1992,proctor1994,pan1999}
Figure~\ref{fig:laboratory} displays an x-ray diffraction curve of the as-grown QW structure
with AlAs/GaAs SL barriers  near the GaAs 004 reflection. The simulation below was performed for a
10~nm thick In$_{0.19}$Ga$_{0.81}$As stressor layer in order to illustrate the source of strain
in the structure. The lattice parameter of the (In,Ga)As stressor layer is obviously larger than
that of the (Al,Ga)As/GaAs QW structure. This is indicated by the smaller Bragg~angle corresponding to
the In$_{0.19}$Ga$_{0.81}$As film. From the positions of the satellite reflections of the AlAs/GaAs
barrier SL, we obtained an average SL period of $\Lambda_0$~=~3.64~nm.
Figure~\ref{fig:PR} shows displacement profiles inside the layer stack obtained by the x-ray  phase retrieval
method.\cite{nikulin1998,nikulin1999} $u$ denotes the displacement and $h$ the magnitude of the reciprocal lattice vector of the corresponding reflection (004). The position of the GaAs QW is marked by an arrow. These profiles were calculated from measurements using a highly intense and highly collimated micro-beam of synchrotron radiation as an incident beam. In this way, a lateral resolution of about 5.0~$\mu$m was achieved during the x-ray diffraction experiment.
Far away from the micro-tube, the displacement profile as expected from the nominal structure of the layer stack was obtained. Even the individual periods of the short period SLs  can be observed. In the vicinity of the micro-tube we detected stress relaxation near the QW and the barrier layers (Fig.~\ref{fig:PR}).  The tube itself could not be measured even in the synchrotron experiment. The very high collimation of the synchrotron beam (quasi-plane wave) leads to rather narrow regions where diffraction conditions are fulfilled in a structure with as small radius of curvature as 20~$\mu$m. This results in a severe reduction of the diffracted signal.

\begin{figure}[!t]
\includegraphics[width=8.0cm]{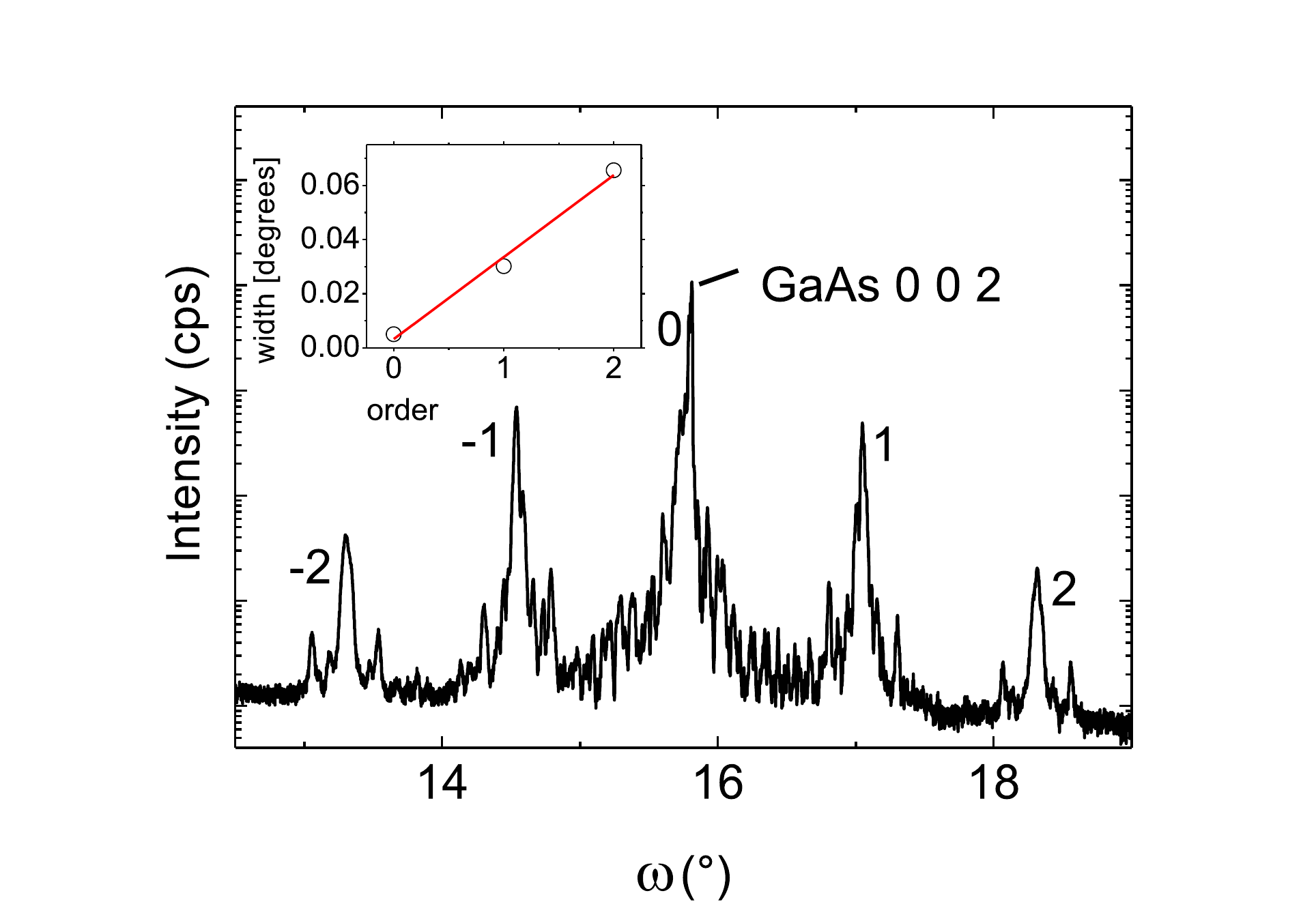}
\caption{  XRD curve of the as-grown (Al,Ga)As/GaAs QW structure with AlAs/GaAs SL barriers and an (In,Ga)As stressor film near the quasi-forbidden GaAs 002 reflection. The satellite maxima of the SLs are marked by their order. The full width at half maximum of the corresponding satellite reflections is plotted in the inset.}
\label{fig:superlattice}
\end{figure}

\begin{figure}[!t]
\includegraphics[width=8.0cm]{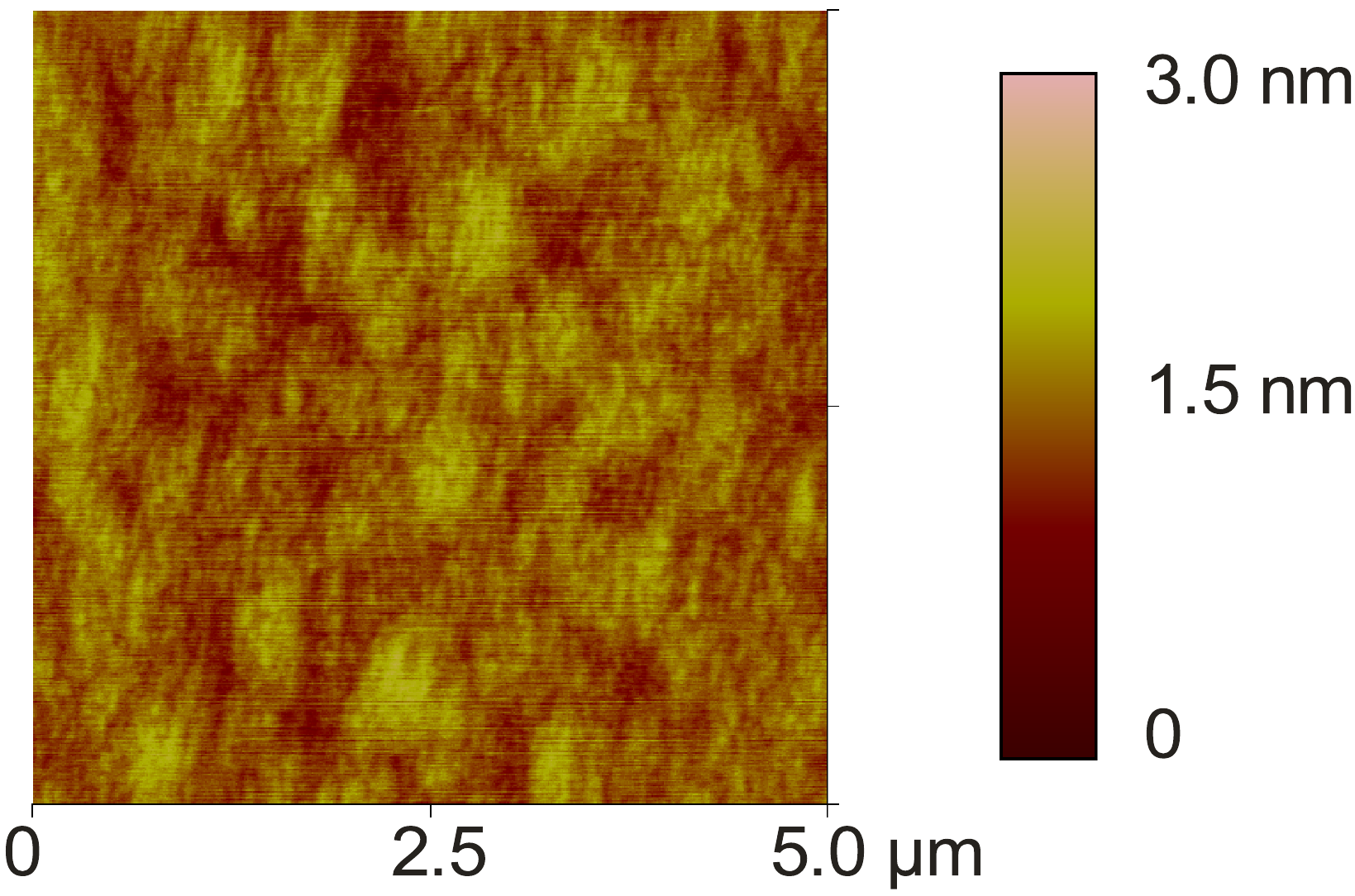}
\caption{ (Color online) AFM micrograph of the sample surface. Besides the extraordinary low value of the RMS roughness of about 0.2~nm, there are surface inhomogeneities on a larger lateral length~scale of about 1.0~$\mu$m. They are caused by step bunching during MBE growth and amount to 2~nm.
A clear azimuthal asymmetry of the island shape is observed.}
\label{fig:AFM}
\end{figure}

\begin{figure}[!t]
\includegraphics[width=12.0cm]{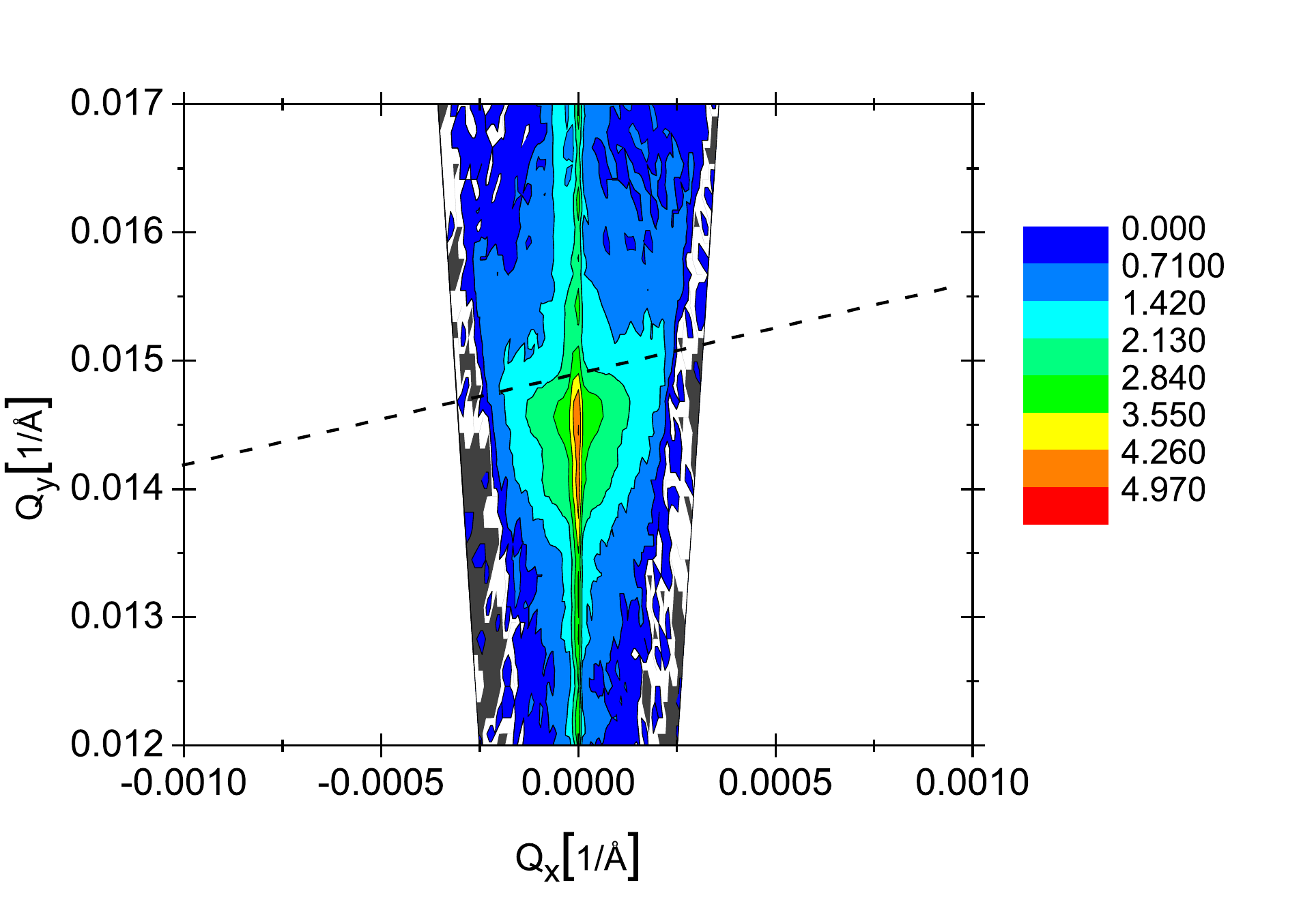}
\caption{ (Color online) Distribution of the x-ray diffuse scattering due to interface roughness of the short period superlattices in the as-grown structure. The diffuse sheet near the first-order diffraction maximum of the short-period superlattices is shown. From the
orientation of this diffuse sheet in reciprocal space (dashed line) an inheritance angle of  about 60$^{\circ}$ can be determined. The intensity scale is logarithmic. The Q$_X$ scaling is stretched for better visibility.}
\label{fig:XRAY}
\end{figure}

The presence of AlAs/GaAs SL barriers in the QW structures opens up the possibility of a direct measurement
of the thickness fluctuation inside the layer stack. Note that, before the QW structure with its
AlAs/GaAs SL barriers was fabricated, a thick GaAs buffer layer was grown in order to start the growth of the QW structures with
a clean surface without substrate defects. During this
buffer layer growth, the substrate roughness was not simply repeated, but a certain modification  by step bunching can be expected, especially
during MBE growth in the step-flow mode. We now characterize the two different kinds of roughness arising in our structures. First we look at the RMS roughness. Neglecting absorption, the diffracted intensity of a SL with $M$ periods can be written as~\cite{speriosu1984}
\begin{equation}\label{Speriosu}
\begin{split}
I = |F|^2 ~\frac{\sin^2(M\Phi)}{\sin^2(\Phi)},
\end{split}
\end{equation}
where
\begin{equation}\label{Speriosu2}
\begin{split}
\Phi(\Lambda)= \frac{\pi\Lambda~\sin~(2{\Theta}_B)}{\lambda|~{\gamma}_H|} [\Theta - {\Theta}_B~+~{\epsilon}^{\bot}~\tan~({\Theta}_B)].
\end{split}
\end{equation}
$\Lambda$ denotes the SL period, ${\Theta}_B$ the Bragg angle of the substrate, and $\Theta$ the
grazing angle of incidence with respect to the diffracting planes, which we assume to be parallel to the surface. $\gamma_H$ is the direction cosine of the diffracted wave,  $\lambda$ the x-ray wavelength, and ${\epsilon}^{\bot}$~the strain  normal to the surface. $F$ is the structure factor of the SL and is a slowly varying function of $\Theta$.
Let $\Phi=n\pi+\Delta$ with $n$ being the peak order.
Around $\Phi=n\pi$, i.e. near a satellite position,  Eq.~(\ref{Speriosu}) can be approximated by a smoothed
function ~\cite{zachariasen1945,pan1999}
\begin{equation}\label{zachariasen}
\begin{split}
I = |F|^2~M^2~\exp~({-M^2\Delta^2/\pi}).
\end{split}
\end{equation}
The full width at half maximum (FWHM) of such a smoothed peak of an ideal SL  can be written as~\cite{zachariasen1945,pan1999}
\begin{equation}\label{FWHM0}
\begin{split}
w_0 = 2\sqrt{\frac{\ln (2)} {\pi}} \frac {\Delta\Theta_M} {M},
\end{split}
\end{equation}
where $\Delta\Theta_M$ denotes the angular spacing between neighboring satellite peaks
\begin{equation}\label{Sdistance}
\begin{split}
\Delta\Theta_M = \frac{\lambda~|\gamma_H|} {\Lambda~\sin{~(2\Theta_B)}}.
\end{split}
\end{equation}
Let us now describe
the period of the real SL as $\Lambda = \Lambda_0 + x$. The fluctuation $x$ of $\Lambda$
can be described by a Gaussian distribution function with standard deviation $\sigma$, where $\Lambda_0$
is the center value of the distribution function and $x$ is the deviation from $\Lambda_0$.
Then, the diffracted intensity becomes
\begin{equation}\label{Ifluctuating}
\begin{split}
I =  \int{|F|^2 \frac {\sin^2[M\Phi(\Lambda_0 + x)] }{\sin^2[\Phi(\Lambda_0 + x)]}[\exp(-\frac{2x^2}{\sigma^2})]^2}dx.
\end{split}
\end{equation}
In this approximation, we can describe the broadening of the $n$-th satellite peak as~\cite{pan1999}
\begin{equation}\label{wfluctuating}
\begin{split}
w_n = w_0 + \sqrt{\ln(2)} \cdot \Delta\Theta_M \cdot n \cdot \frac {\sigma} {\Lambda_0}\;.
\end{split}
\end{equation}
Here, the first term represents an intrinsic width of the satellite peaks, and the second term is a
result of the periodicity fluctuation~$\sigma$. From the experimentally determined peak-widths $w_n$,~the magnitude
of the periodicity fluctuation can be obtained.
Figure~\ref{fig:superlattice} displays an XRD curve near the quasi-forbidden GaAs 002 reflection. The satellite maxima of the AlAs/GaAs SLs forming the barriers of the QW are marked by their order. Higher-order satellites are clearly broadened compared to the zeroth order. An evaluation of the broadening of the satellite maxima using Eq.~(\ref{wfluctuating}) yields a fluctuation value of $\sigma$~=~0.1~nm , which is a measure for the RMS roughness. Similar values for the surface and interface RMS roughness were determined by x-ray reflectivity measurements (not shown here).
Figure~\ref{fig:AFM} shows an AFM micrograph of the surface of the same sample. The values of the RMS interface roughness of $\sigma$~=~0.1~nm determined by XRD
and the RMS surface roughness of about 0.2~nm as determined by AFM are in good agreement. Besides the extraordinary low RMS roughness of about 0.2~nm, there are surface inhomogeneities on the larger lateral length~scale. Their amplitude amounts to 2~nm. A clear azimuthal asymmetry of the island shape is observed.  However, the AFM investigations
showed that besides the small-scale average roughness there are other inhomogeneities of the surface on a larger lateral lengthscale of 1.0~$\mu$m. They are caused by the step-bunching during MBE growth resulting in a clear azimuthal anisotropy of the surface roughness. Such large-scale inhomogeneities are typical for MBE growth in the step-flow growth mode and have already been reported in Refs.~\onlinecite{Kondrashkina1997,jenichen1997}. In addition to the differences between several growth modes, the skew inheritance of the interface roughness during epitaxial growth of semiconductor superlattices was revealed. Figure~\ref{fig:XRAY} demonstrates the intensity distribution of the diffuse x-ray scattering under grazing incidence in order to check the inheritance angles $\alpha$ for our sample. Along GaAs[110], we found an inheritance angle $\alpha_0~\simeq~$60$^{\circ}$, while along the perpendicular direction  [1$\bar{1}$0] it was $\alpha_{90}$~=~83$^{\circ}$.  As a result, we really found a skew inheritance leading to layer thickness inhomogeneities on the larger lateral length scale. The step bunching leads to thickness inhomogeneities in the short period SLs, the QW, and the (In,Ga)As stressor layer.

We will now estimate the effect of the thickness inhomogeneities of the (In,Ga)As stressor layer on the local strain and the
 electronic energy of the structure. Calculations of the strain distribution in the tubes  were carried out by the finite-element method~\cite{elmer2012}. Exact strain calculations for tubes with varying layer thicknesses require complex and time-consuming three-dimensional calculation models. In order to simplify the numerical procedure, calculations were performed in two steps. In the first step, we analytically determined the strain field in a layer stack with a flat (In,Ga)As layer by assuming that it remains flat after detachment from the substrate.
%
%
The thicknesses and elastic constants of the layers in the stack as well as the in-plane strain due to the lattice mismatch ($\epsilon_{\text{epi} }$) relative to the GaAs substrate are given in Tab.~\ref{tab:addlabel}. The strain calculated in this way was used as the driving force for the rolling process.  The stack was allowed to roll along one direction ($x$-direction): the evolution of the strain during the rolling process was determined by a two-dimensional finite element approach.

\begin{figure*}[tbhp]
  \includegraphics[width=0.7\textwidth,angle=0,clip]{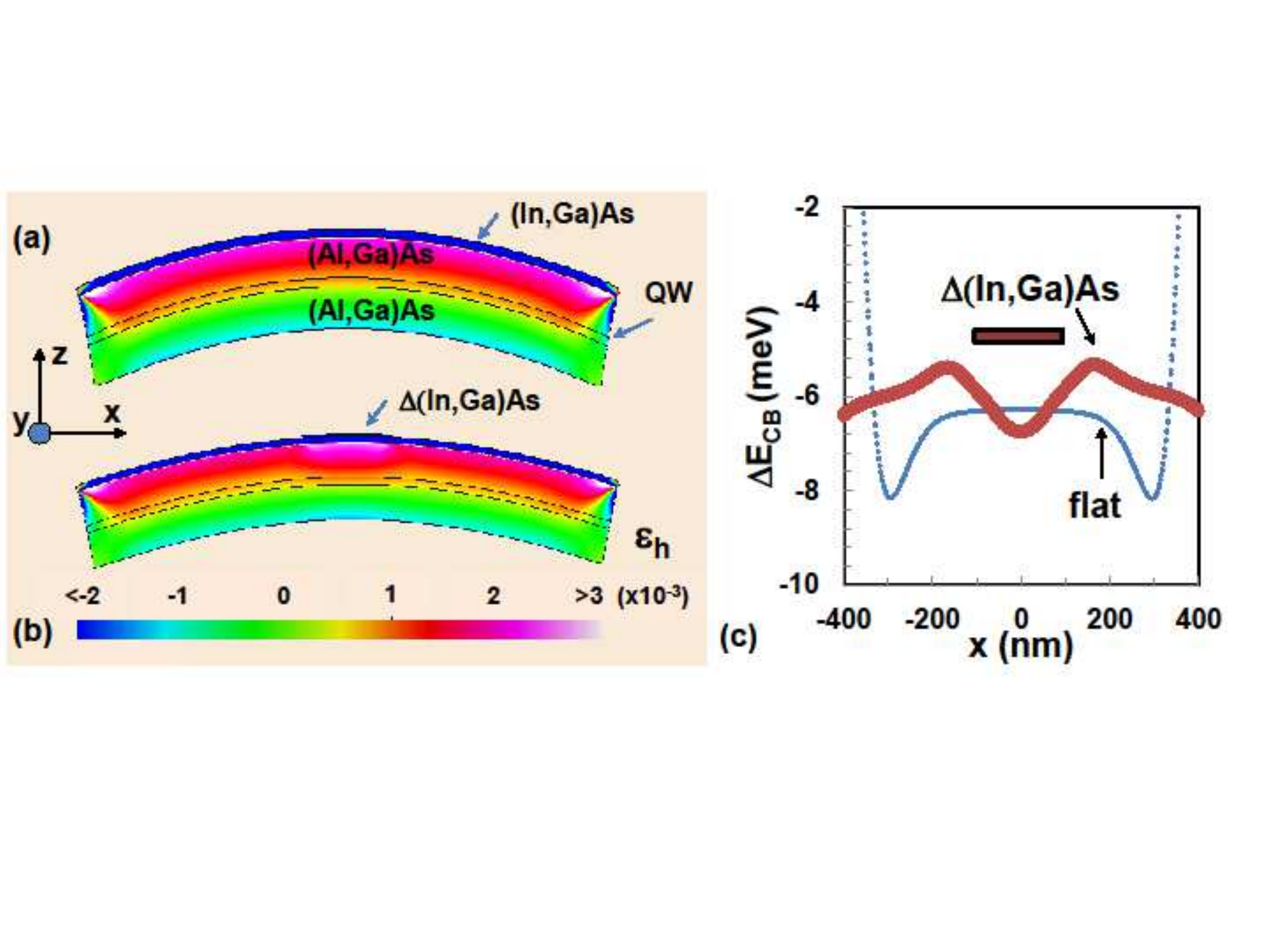}
  \caption{
 \label{calc}
(Color online) Spatial distribution of the hydrostatic strain $\epsilon_h$ in a roll with (a) constant (In,Ga)As thickness (13~nm) and (b) varying (In,Ga)As thickness [thickness of 15~nm in the region indicated as $\Delta$(In,Ga)As and 11~nm elsewhere]. (c) Variation of the conduction band energy $\Delta E_\mathrm{CB}$ along the QW in (a) (thin line) and (b) (thick line).
}

\end{figure*}

Figure~\ref{calc}(a) displays the distribution of the hydrostatic strain $\epsilon_h=\epsilon_{xx}+\epsilon_{yy}$ calculated for a structure with a flat (In,Ga)As stressor (13~nm thick).
The strain variation close to the left and right ends of the roll are due to relaxation. In contrast, the strain in the center region is approximately constant in the QW plane. Note that due to the bending the hydrostatic strain changes from tensile in the upper (Al,Ga)As layer to compressive in the lower one, the plane of zero strain strain (i.e., $\epsilon_h=0$) being located above the GaAs QW.
Figure~\ref{calc}(b) shows the results of similar calculations for a structure with a 11~nm thick (In,Ga)As layer containing a 200~nm long region [indicated by $\Delta$(In,Ga)As] with a larger thickness of 15~nm. The additional tensile strain induced by this region extends to a depth comparable to its width, thereby modifying the strain distribution in the QW.  Figure~\ref{calc}(c) displays profiles for the variation of the conduction band (CB) energy $\Delta E_{\text{CB}}=a_h \epsilon_h$ (relative to the CB energy of  unstrained GaAs) along the QW plane calculated assuming a CB deformation potential $a_{\text{CB}}=-9$~eV~\cite{yu1995}.  A careful examination of the strain profile along the QW plane in Fig.~\ref{calc}(c) shows that $\epsilon_h$ is larger below the $\Delta$(In,Ga)As region. As for the strain distribution in Figs.~\ref{calc}(a) and \ref{calc}(b), the $\Delta E_{\text{CB}}$ variations at the ends of the roll are due to strain relaxation. For the roll with an (In,Ga)As layer of constant thickness,  $\Delta E_{\text{CB}}$ is approximately constant in the central region of the roll. For the roll in Fig.~\ref{calc}(b), the increased thickness of the (In,Ga)As  layer induces an increase in the tensile strain close to the center of the roll, which increases $|\Delta E_{\text{CB}}|$ by approximately 1.6~meV.

\begin{figure}[!t]
\includegraphics[width=10.0cm]{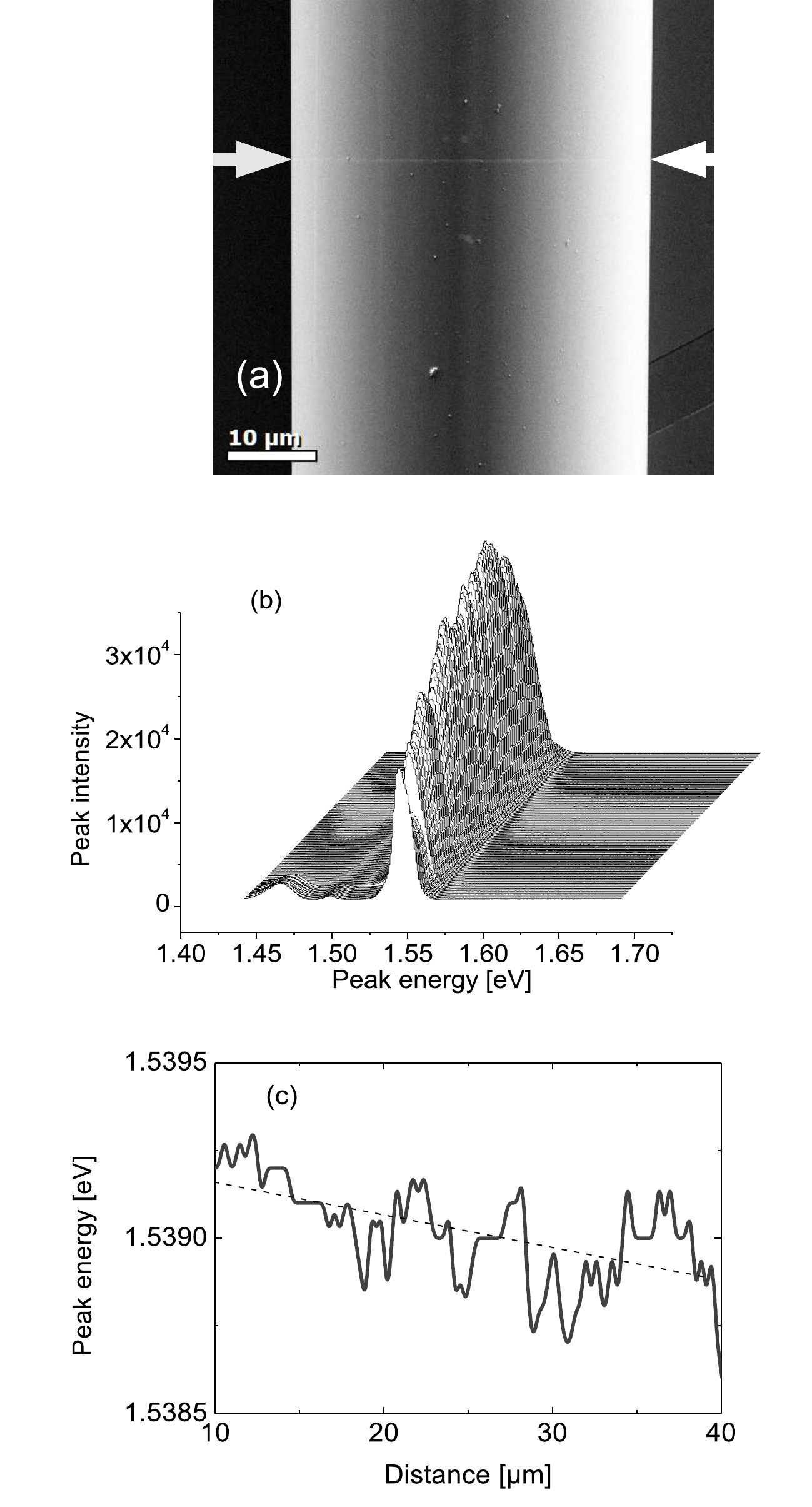}
\caption{(a) The position of a CL line scan performed in the SEM over a typical micro-tube is marked by arrows. (b) displays all the energy spectra corresponding to the line scan shown in (a). The variation of the peak position translates directly to the spatial variation of $\Delta$E$_{CB}$. On (c) the lateral distribution of the peak energies along this CL line scan is plotted. Inhomogeneities of about 0.2~meV on a micrometer length scale are visible. The acceleration voltage was 5~kV while the beam current was 5~nA at a sample temperature of 7~K.}
\label{fig:PeakEnergy}
\end{figure}

A highly sensitive method for the characterization of QW structures with high lateral resolution is CL spectroscopy performed in an SEM.  We applied CL spectroscopy for the characterization of the curved QWs,  prepared from those samples, which were already investigated by XRD and AFM. The electron beam was scanned across a typical micro-tube containing the GaAs QW. A line scan is depicted in in Fig.~\ref{fig:PeakEnergy}(a).  The corresponding energy spectra are given in Fig.~\ref{fig:PeakEnergy}(b). The  CL intensity distribution (not shown here) is very similar to the AFM surface topography pattern in Fig.~\ref{fig:AFM}. This proves that the inhomogeneities of the CL energy are caused by the thickness fluctuation of the GaAs QW in the micro-tube and/or strain fluctuation due to the thickness fluctuation in the AlAs/GaAs SL barriers. The corresponding spatial distribution of the energy detected during the  CL line scan is shown in Fig.~\ref{fig:PeakEnergy} (c). This energy variation directly reflects the variation of the bandgap and at the same time the variation of the chemical potential. The CL line is due to the bandgap luminescence for the given QW with especially high electron density. The fluctuation of the peak energy of the CL line reflects the fluctuation of the chemical potential $\mu_F$ of the QW on a micrometer-scale and amounts to 0.2~meV.

\begin{figure*}[tbhp]
  \includegraphics[width=0.7\textwidth,angle=0,clip]{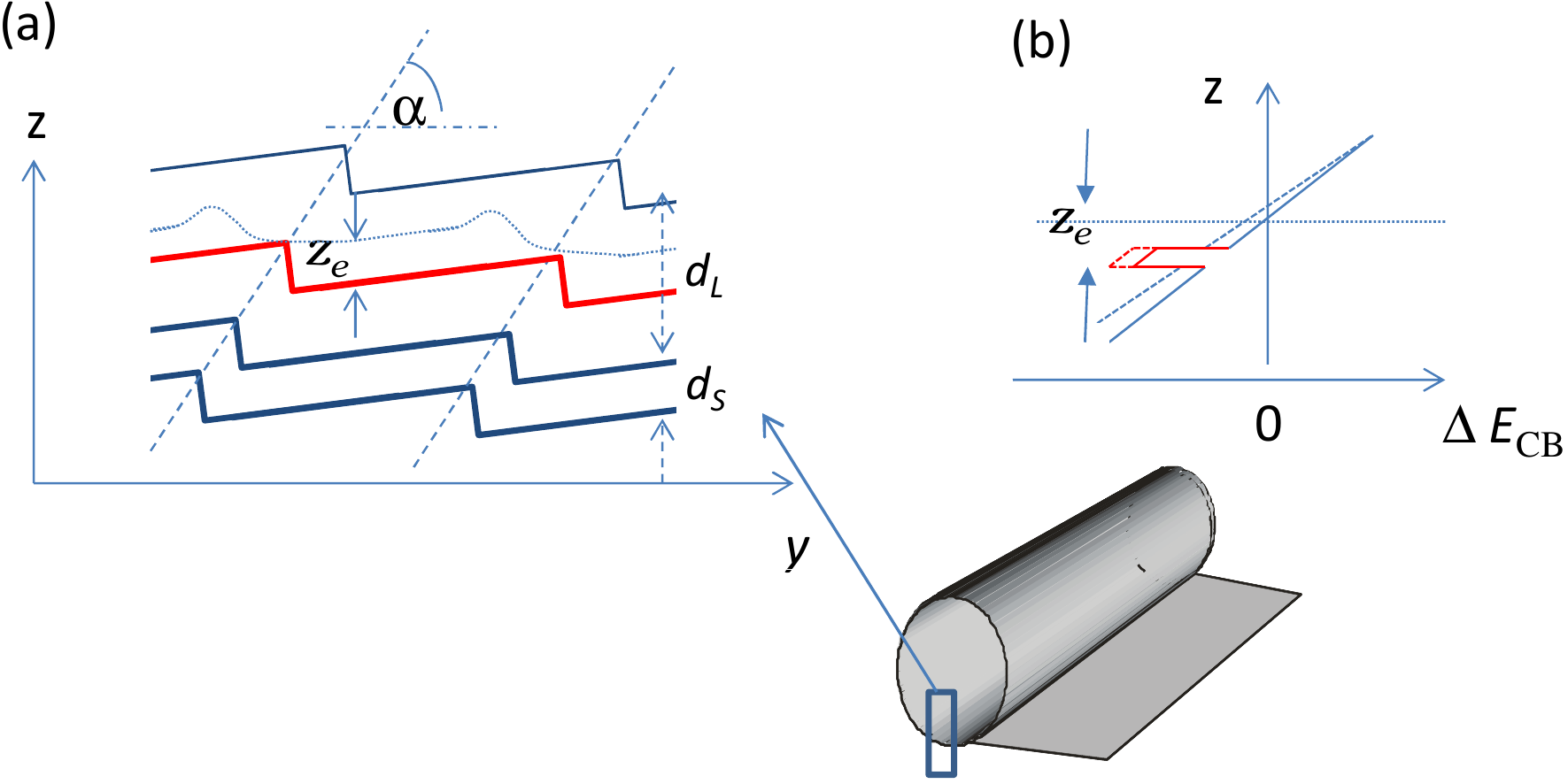}
  \caption{(Color online) Simplified sketch to demonstrate (a) the step inheritance of
the substrate roughness with a finite inclination angle $\alpha$ and (b) the
shift of the conduction band edge $\Delta E_{CB}$  along the (Al,Ga)As
layer stack $z$ coordinate for two different thicknesses $d_S$ of the
(In,Ga)As stressor layer. In (a), the thick blue and red  lines represent
the (In,Ga)As stressor layer  and the 2DEG, respectively. The thin dotted
line represents the zero strain position as estimated from
$d_S(y)$ and the layer thickness and $d_L(y)$.
 \label{fig3}
}

\end{figure*}

Now we can develop a simple model  for the estimation of  density fluctuation in the
given cylindrical structures caused by thickness fluctuation resulting from the step bunching due to an inclined step
height inheritance during the  MBE  growth (cf. Fig.~\ref{fig3}). Note that the QW
as a part of the layer stack has a similar roughness as the structure below,
which is however shifted laterally due to the finite inclination (inheritance) angle $\alpha$.
By releasing and rolling up the layer stack containing the QW (cf. Fig.~\ref{fig:bimetall}), the strain $\Delta_\bot\varepsilon$ along the $z$-direction  normal to the layer arises and fluctuates along the cylinder surface, indicated as the $y$-direction per definition in Fig.~\ref{fig3}, due to variations of the thicknesses $d_{L}$ and $d_{S}$. Consequently, the CB energy $E_{\text{CB}}$  shifts according to the actual value of $\Delta_\bot\varepsilon$.
For our  particular (In,Ga)As/(Al,Ga)As  layer stack,  we calculated the energy shift $\Delta E_{\text{CB}}$  in accordance with the model described in Ref.~\onlinecite{Hey2009}. The results of these calculations are shown in Fig.~\ref{fig3}(b)
for two different thicknesses $d_S$ of the (In,Ga)As stressor layer. The lower part of the layer stack,
which contains the (In,Ga)As stressor layer and also the QW, is under tensile strain.
The upper part of the stack is under compressive strain.
The position of zero~strain and thereby zero-energy~shift  is marked in
Fig.~\ref{fig3}(a) by a dotted line derived from $d_{L}(y)$ and $d_{S}(y)$. Taking this
into account, we assume that  the distance $z_e(y)$ between
the zero-strain line and the position of the 2DEG fluctuates with an amplitude which is
similar to the  layer thickness fluctuation $\delta z_\varepsilon\simeq$ 1~nm.
As a result, we estimate the fluctuation of the conduction band energy
as $\delta E_{\text{CB}}~=~E_\bot~\cdot~\delta z_\varepsilon~\approx~0.3$~meV,
where $ E_\bot=dE_{\text{CB}}/dz = 300$~kV/m denotes the strain induced electrical field. This value is in good
correspondence to the energy fluctuation observed in the CL measurement.

\subsection {Phase cancelation of the SdH oscillations}

\begin{figure}[!b]
\includegraphics[width=8.0cm]{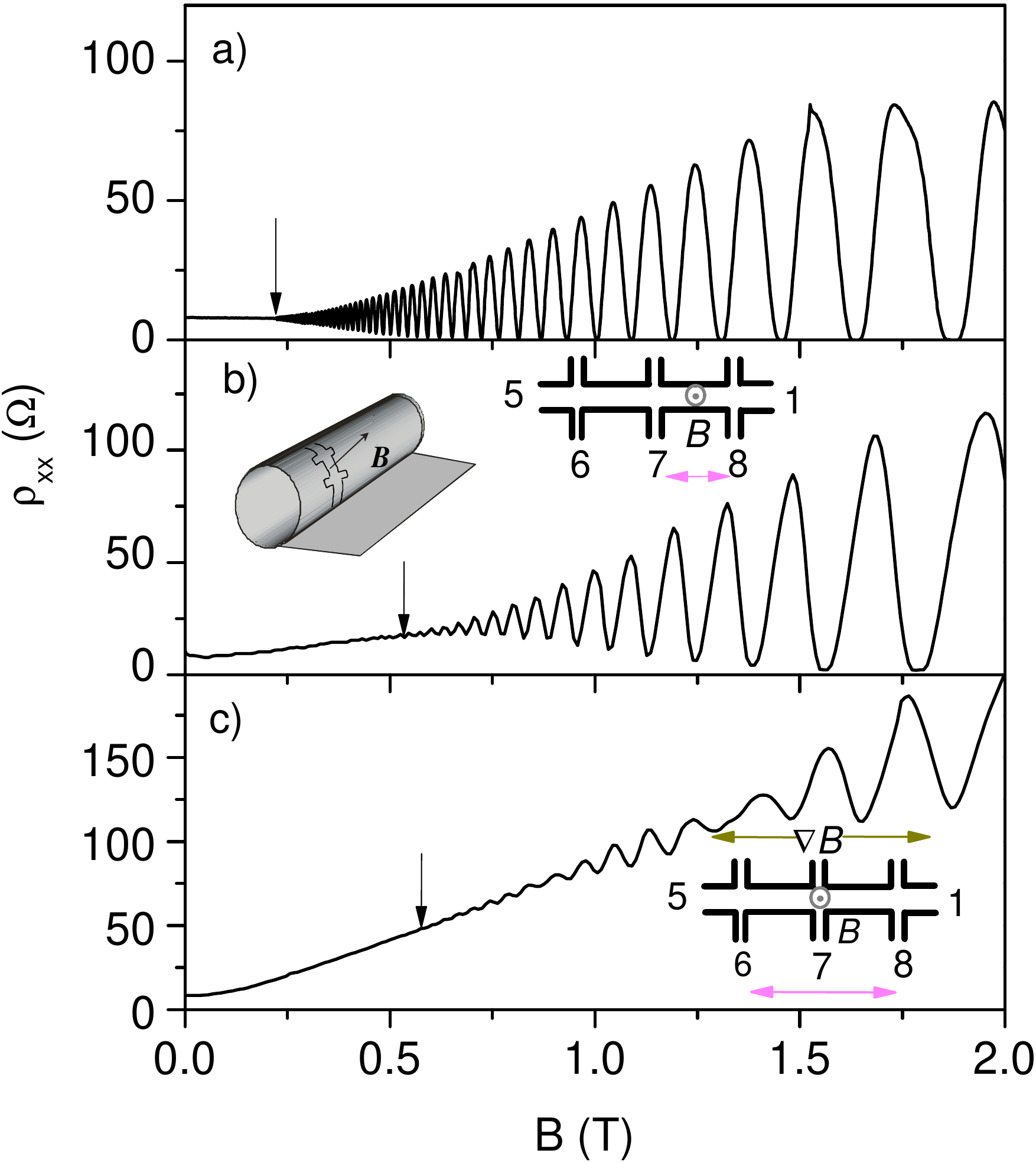}
\caption{ (Color online) Dependence of the resistivity $\rho_{xx}$ on the magnetic field
for Hall bars on the original flat surface (a) and
on cylindrical surfaces (b) and (c).
The insets show  the  Hall
bar geometry and location on the cylinder schematically. The current is imposed
in the leads 1 and  5 , the voltage is measured at leads 7 and 8 (b), 6 and 8 (c),
with a distance of $l_T=10~\mu$m and $l_T=20~\mu$m,  respectively.
The arrows mark the onset of the SdH oscillations. The magnetic
field is oriented perpendicular to the layer for the flat sample and exactly at the position
in the middle between the terminals for resistivity  measurements for the cylindrical sample.
${\nabla \textbf{B}}$ indicates the gradient of the magnetic field. The measurements are performed
 at  $T$~=~50~mK.}
\label{fig1}
\end{figure}

\begin{figure}[]
\includegraphics[width=8.0cm]{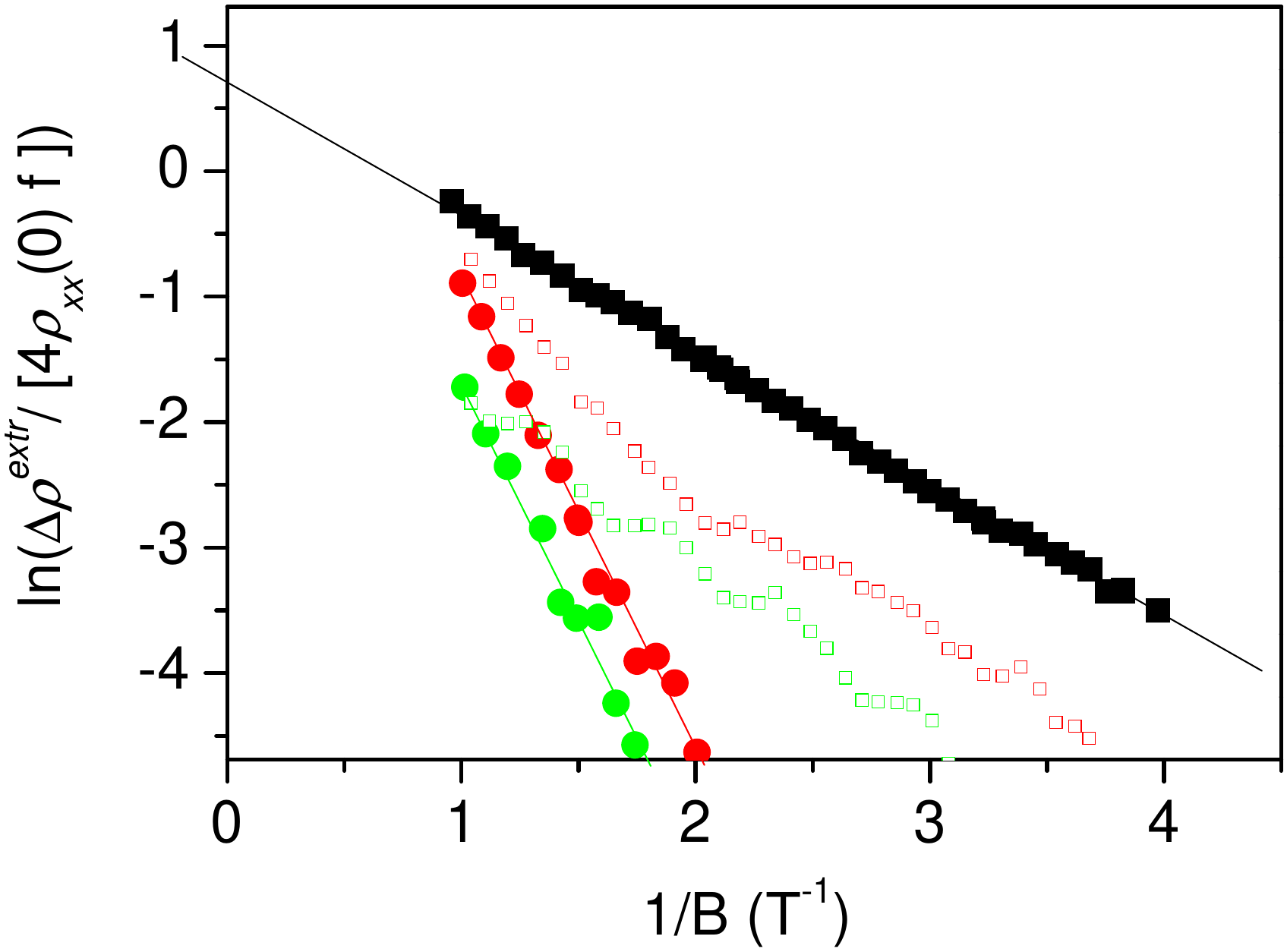}
\caption{ (Color online) Dingle plot for the flat sample (full black squares) and
for the curved sample at terminals with distance $l_T=10~\mu$ (full red circles) and $l_T=20~\mu m$ (full green circles) for measurements shown in Fig.~\ref{fig1}~(a) and (b) respectively. The empty squares represent $\rho_{\text{aver}}$ the    averaged
 flat sample data  along  the corresponding magnetic field gradients in correspondence with  measurements shown in Fig.~\ref{fig1}~(a) and (b).}
\label{fig2}
\end{figure}

Figure~\ref{fig1} compares the longitudinal resistivities $\rho_{xx}$ as a function of the magnetic field $B$ in the region of the SdH oscillations for Hall bar structures  on flat (a) and cylindrical surfaces (b) and (c).
The terminals for the measurement of $\rho_{xx}$ are separated by $l=10~\mu$m and $l=20~\mu$m
for the traces in Figs.~\ref{fig1}~(b) and
(c), respectively.
In Tab.~\ref{table1} the results of the cancelation of the SdH oscillations are collected. Note that the low-field mobilities and
carrier densities do not differ significantly in both samples. This shows
that the self-rolling process does not introduce additional defects and surface depletion
of the HM2DEG density. However, the magnitude of the SdH oscillations is strongly  reduced in the
cylindrical structures. Moreover, visible  SdH oscillations
appear at much higher magnetic field values of about $B_{\text{onset}}$~=~0.5~--~0.6~T for the
cylindrical surface  as compared to  0.2 T for the flat surface.
The so-called 'Dingle plot' is   used to characterize  the magnetic field
dependence of the magnitude of the  SdH oscillations in $\rho_{xx}$.\cite{Coleridge1991}
It allows for an estimation of the total scattering time $\tau _{\text{tot}}$  from $\Delta \rho^{\text{extr}}$, which is
the difference between the zero-field resistance $\rho_{xx}(0)$ and $\rho^{\text{extr}}_{xx}$ at the resistance
oscillation extremes in accordance with the conventional Ando formula
\begin{equation}\label{Ando}
\begin{split}
\frac {\Delta \rho^{\text{extr}}}  {\rho_{xx}(0)}= f 4 e^{-\pi/(\omega _c \tau _{\text{tot}})}~,
\end{split}
\end{equation}
where $f=A_T/\sinh(A_T)$ with $A_T=2\pi ^2 k_BT/(\hbar \omega_ c$) accounts for thermal smearing.
 $\omega_ c=eB_{\text{tot}} /m^*$ denotes the cyclotron frequency and ~$m^*$
the  effective electron mass.\cite{Mancoff1996}
Fig.~\ref{fig2} demonstrates the Dingle-plots for the flat sample (a) and for the curved samples (b) and (c).
These graphs present the function ln($\Delta \rho^{\text{extr}}/[4\rho_{xx}(0)~f$]) plotted versus $1/B$. For the flat sample, the experimental
points follow precisely the Ando formula. They can be approximated by a straight line, which intersects the axis $1/B$ = 0
at the ordinate value of $1$. We can estimate a total scattering time  $\tau _{\text{tot}}$ = 1.12 ps, which is mostly
governed by small-angle scattering processes, in
contrast to the much larger transport scattering time $\tau _{\text{tr}}$ = 50 ps, which
is determined by large-angle scattering events
 (cf. Tab.~\ref{table1}).

The strong reduction of the SdH oscillations in curved samples increases the slope of the corresponding lines in this
representation, which leads to nearly three times lower
values of $\tau _{\text{tot}}$ as compared to the flat one (cf. Tab.~\ref{table1}). In addition, the extrapolation
of the linear fit does not intersect the $1/B$ = 0 axis at the ordinate value of $1$.
Such a behavior was earlier discussed
in the context of sample inhomogeneities.\cite{Coleridge1991} An obvious
reason for sample inhomogeneities  is the curvature of the Hall bar
resulting in a magnetic field gradient, which surely leads to a smearing of the SdH oscillation.

In order to prove, if the smearing effect due to the magnetic field gradient can explain the low  values of $\tau _{\text{tot}}$, we present in Fig.~\ref{fig2} also the results of averaging the data of the  flat sample $\rho_{xx}^{\text{flat}}$  [cf. Fig.~\ref{fig1}(a)].
We calculate the average value  $\rho_{\text{aver}} = \delta l /l_T \sum \rho_{xx}^{\text{flat}}[B_0$ cos($\phi_i)$]. Here, sin($\phi_i$)=$i\delta l/r$, $\delta l=0.1~\mu$m, $i$ = 0,1,2,.. up to $i\delta l =l_T$, for the terminal distances
$l_T = 10~\mu$m and $l_T = 20~\mu$m according to the measurements [cf. Figs.~\ref{fig1}(b) and~\ref{fig1}(c)].  The corresponding Dingle-plots
 show oscillations, which arise from the beating of two modes, belonging to the
 minimum and maximum values of the magnetic field in the Hall bar. Such a
 beating was already observed in the SdH oscillations at much larger gradients of the field.\cite{Friedland2009}
 In fact, this averaging by curvature  reduces the magnitude of the SdH oscillations effectively, but it does not  change the slope of the Dingle plot.  Moreover, this  effect
 does not eliminate the large extrapolation values at $1/B$ = 0. It seems that the
 experimental data can be approximated by a straight line corresponding to the averaging by means of the magnetic field
 gradient at higher magnetic fields. Unfortunately, all the data at magnetic fields
 above 1~T show a clear signature of the quantum Hall effect, which results in a completely different
 dependence of $\rho_{xx}$ on the magnetic field.\cite{Friedland2009}

 Another source of lateral inhomogeneities may be the so-called 'static skin'
 effect.\cite{Chaplik2000,Friedland2010} The gradient of the field results in a
 redistribution of the current flow toward one of the edges of the Hall bar, which
 changes the side by reversing the magnetic field direction.  For the present geometry,
 the  skin length $l_{\text{skin}} = 1/\mu\Delta B$, which determines the width of the
 conducting channel, is below 1~$\mu$m at the position with the maximal
 field gradient   $\nabla \textbf{B} \sim \sin (l_T/R) $T.  $\nabla \textbf{B}$ is twice as large in the
 measurement shown in Fig.~\ref{fig1}(c) compared to measurement shown in Fig.~\ref{fig1}(b). However, both samples
 behave nearly identically in the Dingle plots, so that we have to reject the static skin effect as a cause
 for the reduced SdH oscillations in the curved samples.

Finally, we discuss the consequences of  strain fluctuation.
In a cylindrical structure with a HM2DEG, the resulting energy fluctuation is a fluctuation of the conduction band energy. Therefore, the chemical potential $\mu_F$ and the carrier density  $N$ of the HM2DEG vary in accordance with the strain fluctuation.   The impact of density fluctuation on the magnitude of the SdH oscillations was already identified as a phase cancelation, which  reduces the
 magnitude of the SdH oscillation due to a broadening of  the  Landau~level
width. The corresponding change in density necessary for the phase~cancelation of the
 SdH oscillations is given in Ref.~\onlinecite{Harrang1985} as
\begin{equation}\label{flucDens}
\begin{split}
\delta N = \frac{eB_{\text{onset}}} {4\pi\hbar} ~.
\end{split}
\end{equation}
With the observed onset $B_{\text{onset}}\cong$~0.56 T of the SdH  oscillations in the cylindrical structures, we estimate  $\delta N~=~0.7\times10^{14}$~m$^{-2}$. This value may be used to calculate
the fluctuation $\delta \mu_F\simeq 0.22$~meV  and $\delta k_F/k_F\simeq 10^{-2}$
of the chemical potential and the relative Fermi wave vector, respectively. The
latter introduces a scattering angle of about one degree, which proves that the observed phase cancelation is a low-angle
scattering effect. We estimate the quantum relaxation time due to density
fluctuation $\tau_{\delta N} = \hbar/(2 \delta \mu_F) \simeq$ 1.4~ps. Then, the
value $(1/\tau_{\delta N}+1/\tau^{\text{flat}}_{\text{tot}})^{-1} \simeq$ 0.64~ps is in a good
agreement with the measured low-angle scattering time $\tau_{\text{tot}}$ in the cylindrical
structures (cf. Tab.~\ref{table1}). The value of $\delta \mu_F$ is in good agreement with $\Delta\textit{E}$ obtained from the energy
fluctuation of the QW obtained by CL spectroscopy.
We can estimate the density
fluctuation  as $\delta N \simeq (0.9-1.8) \times10^{14}~$ m$^{-2}$ from the fluctuation of the conduction band energy $\delta E_c$,
which is in a good agreement with the  value estimated from the  damping of the SdH oscillations.\newline

\section{Conclusion}
The strain state in released (In,Ga)As/(Al,Ga)As micro-tubes was characterized by strain fluctuation with a lateral correlation length of about 1.0~$\mu$m. This strain fluctuation images the thickness inhomogeneities, which are due to step bunching during epitaxial growth of the original layer. They lead to a local variation of the chemical potential and, therefore, to  fluctuation of the carrier density.  This results in a phase cancelation of the Shubnikov-de-Haas oscillations in the curved HM2DEG and, therefore, in a degradation of the single-electron scattering time, whereas the electron mobility in these structures remains high. The estimated fluctuation of the carrier density agrees well with the energy fluctuation measured in cathodoluminescence spectra of the free-electron transition of the quantum well.

\section{Acknowledgement}
The authors thank A.~Riedel and M.~H\"oricke for sample preparation, H.~T.~Grahn and J. Herfort for a critical reading of the manuscript,
and V.~M.~Kaganer for helpful discussions. We acknowledge the help of O.~Sakata of on BL13XU during the experiment performed at SPring 8.

\section{References}


\begin{thebibliography}{30}%
\makeatletter
\providecommand \@ifxundefined [1]{%
 \@ifx{#1\undefined}
}%
\providecommand \@ifnum [1]{%
 \ifnum #1\expandafter \@firstoftwo
 \else \expandafter \@secondoftwo
 \fi
}%
\providecommand \@ifx [1]{%
 \ifx #1\expandafter \@firstoftwo
 \else \expandafter \@secondoftwo
 \fi
}%
\providecommand \natexlab [1]{#1}%
\providecommand \enquote  [1]{``#1''}%
\providecommand \bibnamefont  [1]{#1}%
\providecommand \bibfnamefont [1]{#1}%
\providecommand \citenamefont [1]{#1}%
\providecommand \href@noop [0]{\@secondoftwo}%
\providecommand \href [0]{\begingroup \@sanitize@url \@href}%
\providecommand \@href[1]{\@@startlink{#1}\@@href}%
\providecommand \@@href[1]{\endgroup#1\@@endlink}%
\providecommand \@sanitize@url [0]{\catcode `\\12\catcode `\$12\catcode
  `\&12\catcode `\#12\catcode `\^12\catcode `\_12\catcode `\%12\relax}%
\providecommand \@@startlink[1]{}%
\providecommand \@@endlink[0]{}%
\providecommand \url  [0]{\begingroup\@sanitize@url \@url }%
\providecommand \@url [1]{\endgroup\@href {#1}{\urlprefix }}%
\providecommand \urlprefix  [0]{URL }%
\providecommand \Eprint [0]{\href }%
\providecommand \doibase [0]{http://dx.doi.org/}%
\providecommand \selectlanguage [0]{\@gobble}%
\providecommand \bibinfo  [0]{\@secondoftwo}%
\providecommand \bibfield  [0]{\@secondoftwo}%
\providecommand \translation [1]{[#1]}%
\providecommand \BibitemOpen [0]{}%
\providecommand \bibitemStop [0]{}%
\providecommand \bibitemNoStop [0]{.\EOS\space}%
\providecommand \EOS [0]{\spacefactor3000\relax}%
\providecommand \BibitemShut  [1]{\csname bibitem#1\endcsname}%
\let\auto@bib@innerbib\@empty
\bibitem [{\citenamefont {Prinz}\ \emph {et~al.}(2000)\citenamefont {Prinz},
  \citenamefont {Seleznev}, \citenamefont {Gutakovsky}, \citenamefont
  {Chehovskiy}, \citenamefont {Preobrazhenskii}, \citenamefont {Putyato},\ and\
  \citenamefont {Gavrilova}}]{prinz00}%
  \BibitemOpen
  \bibfield  {author} {\bibinfo {author} {\bibfnamefont {V.~Y.}\ \bibnamefont
  {Prinz}}, \bibinfo {author} {\bibfnamefont {V.~A.}\ \bibnamefont {Seleznev}},
  \bibinfo {author} {\bibfnamefont {A.~K.}\ \bibnamefont {Gutakovsky}},
  \bibinfo {author} {\bibfnamefont {A.~V.}\ \bibnamefont {Chehovskiy}},
  \bibinfo {author} {\bibfnamefont {V.~V.}\ \bibnamefont {Preobrazhenskii}},
  \bibinfo {author} {\bibfnamefont {M.~A.}\ \bibnamefont {Putyato}}, \ and\
  \bibinfo {author} {\bibfnamefont {T.~A.}\ \bibnamefont {Gavrilova}},\
  }\href@noop {} {\bibfield  {journal} {\bibinfo  {journal} {Physica E
  (Amsterdam)}\ }\textbf {\bibinfo {volume} {6}},\ \bibinfo {pages} {828}
  (\bibinfo {year} {2000})}\BibitemShut {NoStop}%
\bibitem [{\citenamefont {Friedland}\ \emph {et~al.}(2007)\citenamefont
  {Friedland}, \citenamefont {Hey}, \citenamefont {Kostial}, \citenamefont
  {Riedel},\ and\ \citenamefont {Ploog}}]{friedland2007}%
  \BibitemOpen
  \bibfield  {author} {\bibinfo {author} {\bibfnamefont {K.~J.}\ \bibnamefont
  {Friedland}}, \bibinfo {author} {\bibfnamefont {R.}~\bibnamefont {Hey}},
  \bibinfo {author} {\bibfnamefont {H.}~\bibnamefont {Kostial}}, \bibinfo
  {author} {\bibfnamefont {A.}~\bibnamefont {Riedel}}, \ and\ \bibinfo {author}
  {\bibfnamefont {K.~H.}\ \bibnamefont {Ploog}},\ }\href@noop {} {\bibfield
  {journal} {\bibinfo  {journal} {Phys. Rev. B}\ }\textbf {\bibinfo {volume}
  {75}},\ \bibinfo {pages} {045347} (\bibinfo {year} {2007})}\BibitemShut
  {NoStop}%
\bibitem [{\citenamefont {Vorobev}\ \emph {et~al.}(2007)\citenamefont
  {Vorobev}, \citenamefont {Friedland}, \citenamefont {Kostial}, \citenamefont
  {Hey}, \citenamefont {Jahn}, \citenamefont {Wiebicke}, \citenamefont
  {Yukecheva},\ and\ \citenamefont {Prinz}}]{vorobev2007}%
  \BibitemOpen
  \bibfield  {author} {\bibinfo {author} {\bibfnamefont {A.~B.}\ \bibnamefont
  {Vorobev}}, \bibinfo {author} {\bibfnamefont {K.~J.}\ \bibnamefont
  {Friedland}}, \bibinfo {author} {\bibfnamefont {H.}~\bibnamefont {Kostial}},
  \bibinfo {author} {\bibfnamefont {R.}~\bibnamefont {Hey}}, \bibinfo {author}
  {\bibfnamefont {U.}~\bibnamefont {Jahn}}, \bibinfo {author} {\bibfnamefont
  {E.}~\bibnamefont {Wiebicke}}, \bibinfo {author} {\bibfnamefont {J.~S.}\
  \bibnamefont {Yukecheva}}, \ and\ \bibinfo {author} {\bibfnamefont {V.~Y.}\
  \bibnamefont {Prinz}},\ }\href@noop {} {\bibfield  {journal} {\bibinfo
  {journal} {Phys. Rev. B}\ }\textbf {\bibinfo {volume} {75}},\ \bibinfo
  {pages} {205309} (\bibinfo {year} {2007})}\BibitemShut {NoStop}%
\bibitem [{\citenamefont {Friedland}\ \emph {et~al.}(2009)\citenamefont
  {Friedland}, \citenamefont {Siddiki}, \citenamefont {Hey}, \citenamefont
  {Kostial}, \citenamefont {Riedel},\ and\ \citenamefont
  {Maude}}]{Friedland2009}%
  \BibitemOpen
  \bibfield  {author} {\bibinfo {author} {\bibfnamefont {K.~J.}\ \bibnamefont
  {Friedland}}, \bibinfo {author} {\bibfnamefont {A.}~\bibnamefont {Siddiki}},
  \bibinfo {author} {\bibfnamefont {R.}~\bibnamefont {Hey}}, \bibinfo {author}
  {\bibfnamefont {H.}~\bibnamefont {Kostial}}, \bibinfo {author} {\bibfnamefont
  {A.}~\bibnamefont {Riedel}}, \ and\ \bibinfo {author} {\bibfnamefont {D.~K.}\
  \bibnamefont {Maude}},\ }\href@noop {} {\bibfield  {journal} {\bibinfo
  {journal} {Phys. Rev. B}\ }\textbf {\bibinfo {volume} {79}},\ \bibinfo
  {pages} {125320} (\bibinfo {year} {2009})}\BibitemShut {NoStop}%
\bibitem [{\citenamefont {Krause}\ \emph {et~al.}(2006)\citenamefont {Krause},
  \citenamefont {Mocuta}, \citenamefont {Metzger}, \citenamefont {Deneke},\
  and\ \citenamefont {Schmidt}}]{Krause2006}%
  \BibitemOpen
  \bibfield  {author} {\bibinfo {author} {\bibfnamefont {B.}~\bibnamefont
  {Krause}}, \bibinfo {author} {\bibfnamefont {C.}~\bibnamefont {Mocuta}},
  \bibinfo {author} {\bibfnamefont {T.~H.}\ \bibnamefont {Metzger}}, \bibinfo
  {author} {\bibfnamefont {C.}~\bibnamefont {Deneke}}, \ and\ \bibinfo {author}
  {\bibfnamefont {O.~G.}\ \bibnamefont {Schmidt}},\ }\href@noop {} {\bibfield
  {journal} {\bibinfo  {journal} {Phys. Rev. Lett.}\ }\textbf {\bibinfo
  {volume} {96}},\ \bibinfo {pages} {165502} (\bibinfo {year}
  {2006})}\BibitemShut {NoStop}%
\bibitem [{\citenamefont {Malachias}\ \emph {et~al.}(2009)\citenamefont
  {Malachias}, \citenamefont {Deneke}, \citenamefont {Krause}, \citenamefont
  {Mocuta}, \citenamefont {Kiravittaya}, \citenamefont {Metzger},\ and\
  \citenamefont {Schmidt}}]{Malachias2009}%
  \BibitemOpen
  \bibfield  {author} {\bibinfo {author} {\bibfnamefont {A.}~\bibnamefont
  {Malachias}}, \bibinfo {author} {\bibfnamefont {C.}~\bibnamefont {Deneke}},
  \bibinfo {author} {\bibfnamefont {B.}~\bibnamefont {Krause}}, \bibinfo
  {author} {\bibfnamefont {C.}~\bibnamefont {Mocuta}}, \bibinfo {author}
  {\bibfnamefont {S.}~\bibnamefont {Kiravittaya}}, \bibinfo {author}
  {\bibfnamefont {T.~H.}\ \bibnamefont {Metzger}}, \ and\ \bibinfo {author}
  {\bibfnamefont {O.~G.}\ \bibnamefont {Schmidt}},\ }\href@noop {} {\bibfield
  {journal} {\bibinfo  {journal} {Phys. Rev.B}\ }\textbf {\bibinfo {volume}
  {79}},\ \bibinfo {pages} {035301} (\bibinfo {year} {2009})}\BibitemShut
  {NoStop}%
\bibitem [{\citenamefont {Deneke}\ \emph {et~al.}(2009)\citenamefont {Deneke},
  \citenamefont {Songmuang}, \citenamefont {Jin-Phillip},\ and\ \citenamefont
  {Schmidt}}]{Deneke2009}%
  \BibitemOpen
  \bibfield  {author} {\bibinfo {author} {\bibfnamefont {C.}~\bibnamefont
  {Deneke}}, \bibinfo {author} {\bibfnamefont {R.}~\bibnamefont {Songmuang}},
  \bibinfo {author} {\bibfnamefont {N.~Y.}\ \bibnamefont {Jin-Phillip}}, \ and\
  \bibinfo {author} {\bibfnamefont {O.~G.}\ \bibnamefont {Schmidt}},\
  }\href@noop {} {\bibfield  {journal} {\bibinfo  {journal} {J. Phys. D: Appl.
  Phys.}\ }\textbf {\bibinfo {volume} {42}},\ \bibinfo {pages} {103001}
  (\bibinfo {year} {2009})}\BibitemShut {NoStop}%
\bibitem [{\citenamefont {Grundmann}(2003)}]{Grundmann2003}%
  \BibitemOpen
  \bibfield  {author} {\bibinfo {author} {\bibfnamefont {M.}~\bibnamefont
  {Grundmann}},\ }\href@noop {} {\bibfield  {journal} {\bibinfo  {journal}
  {Appl.~Phys.~Lett.}\ }\textbf {\bibinfo {volume} {83}},\ \bibinfo {pages}
  {2444} (\bibinfo {year} {2003})}\BibitemShut {NoStop}%
\bibitem [{\citenamefont {Kondrashkina}\ \emph {et~al.}(1997)\citenamefont
  {Kondrashkina}, \citenamefont {Stepanov}, \citenamefont {Opitz},
  \citenamefont {Schmidbauer}, \citenamefont {K\"ohler}, \citenamefont {Hey},
  \citenamefont {Wassermeier},\ and\ \citenamefont
  {Novikov}}]{Kondrashkina1997}%
  \BibitemOpen
  \bibfield  {author} {\bibinfo {author} {\bibfnamefont {E.~A.}\ \bibnamefont
  {Kondrashkina}}, \bibinfo {author} {\bibfnamefont {S.~A.}\ \bibnamefont
  {Stepanov}}, \bibinfo {author} {\bibfnamefont {R.}~\bibnamefont {Opitz}},
  \bibinfo {author} {\bibfnamefont {M.}~\bibnamefont {Schmidbauer}}, \bibinfo
  {author} {\bibfnamefont {R.}~\bibnamefont {K\"ohler}}, \bibinfo {author}
  {\bibfnamefont {R.}~\bibnamefont {Hey}}, \bibinfo {author} {\bibfnamefont
  {M.}~\bibnamefont {Wassermeier}}, \ and\ \bibinfo {author} {\bibfnamefont
  {D.~V.}\ \bibnamefont {Novikov}},\ }\href@noop {} {\bibfield  {journal}
  {\bibinfo  {journal} {Phys. Rev. B}\ }\textbf {\bibinfo {volume} {56}},\
  \bibinfo {pages} {10469} (\bibinfo {year} {1997})}\BibitemShut {NoStop}%
\bibitem [{\citenamefont {Jenichen}\ \emph {et~al.}(1997)\citenamefont
  {Jenichen}, \citenamefont {Hey}, \citenamefont {Wassermeier},\ and\
  \citenamefont {Ploog}}]{jenichen1997}%
  \BibitemOpen
  \bibfield  {author} {\bibinfo {author} {\bibfnamefont {B.}~\bibnamefont
  {Jenichen}}, \bibinfo {author} {\bibfnamefont {R.}~\bibnamefont {Hey}},
  \bibinfo {author} {\bibfnamefont {M.}~\bibnamefont {Wassermeier}}, \ and\
  \bibinfo {author} {\bibfnamefont {K.}~\bibnamefont {Ploog}},\ }\href@noop {}
  {\bibfield  {journal} {\bibinfo  {journal} {Il Nuovo Cimento D}\ }\textbf
  {\bibinfo {volume} {19}},\ \bibinfo {pages} {429} (\bibinfo {year}
  {1997})}\BibitemShut {NoStop}%
\bibitem [{\citenamefont {Jahn}\ \emph {et~al.}(1994)\citenamefont {Jahn},
  \citenamefont {Fujiwara}, \citenamefont {Menniger}, \citenamefont {Hey},\
  and\ \citenamefont {Grahn}}]{jahn1994}%
  \BibitemOpen
  \bibfield  {author} {\bibinfo {author} {\bibfnamefont {U.}~\bibnamefont
  {Jahn}}, \bibinfo {author} {\bibfnamefont {K.}~\bibnamefont {Fujiwara}},
  \bibinfo {author} {\bibfnamefont {J.}~\bibnamefont {Menniger}}, \bibinfo
  {author} {\bibfnamefont {R.}~\bibnamefont {Hey}}, \ and\ \bibinfo {author}
  {\bibfnamefont {H.~T.}\ \bibnamefont {Grahn}},\ }\href@noop {} {\bibfield
  {journal} {\bibinfo  {journal} {J. Appl. Phys.}\ }\textbf {\bibinfo {volume}
  {77}},\ \bibinfo {pages} {1211} (\bibinfo {year} {1994})}\BibitemShut
  {NoStop}%
\bibitem [{\citenamefont {Jahn}\ \emph {et~al.}(1995)\citenamefont {Jahn},
  \citenamefont {Fujiwara}, \citenamefont {Hey}, \citenamefont {Kastrup},
  \citenamefont {Grahn},\ and\ \citenamefont {Menniger}}]{jahn1995}%
  \BibitemOpen
  \bibfield  {author} {\bibinfo {author} {\bibfnamefont {U.}~\bibnamefont
  {Jahn}}, \bibinfo {author} {\bibfnamefont {K.}~\bibnamefont {Fujiwara}},
  \bibinfo {author} {\bibfnamefont {R.}~\bibnamefont {Hey}}, \bibinfo {author}
  {\bibfnamefont {J.}~\bibnamefont {Kastrup}}, \bibinfo {author} {\bibfnamefont
  {H.~T.}\ \bibnamefont {Grahn}}, \ and\ \bibinfo {author} {\bibfnamefont
  {J.}~\bibnamefont {Menniger}},\ }\href@noop {} {\bibfield  {journal}
  {\bibinfo  {journal} {J. Cryst. Growth}\ }\textbf {\bibinfo {volume} {150}},\
  \bibinfo {pages} {43} (\bibinfo {year} {1995})}\BibitemShut {NoStop}%
\bibitem [{\citenamefont {Jahn}\ \emph {et~al.}(1996)\citenamefont {Jahn},
  \citenamefont {Menniger}, \citenamefont {Hey}, \citenamefont {Jenichen},
  \citenamefont {Runge},\ and\ \citenamefont {Grahn}}]{jahn1996}%
  \BibitemOpen
  \bibfield  {author} {\bibinfo {author} {\bibfnamefont {U.}~\bibnamefont
  {Jahn}}, \bibinfo {author} {\bibfnamefont {J.}~\bibnamefont {Menniger}},
  \bibinfo {author} {\bibfnamefont {R.}~\bibnamefont {Hey}}, \bibinfo {author}
  {\bibfnamefont {B.}~\bibnamefont {Jenichen}}, \bibinfo {author}
  {\bibfnamefont {E.}~\bibnamefont {Runge}}, \ and\ \bibinfo {author}
  {\bibfnamefont {H.~T.}\ \bibnamefont {Grahn}},\ }\href@noop {} {\bibfield
  {journal} {\bibinfo  {journal} {Mater. Sci. Eng. B}\ }\textbf {\bibinfo
  {volume} {42}},\ \bibinfo {pages} {133} (\bibinfo {year} {1996})}\BibitemShut
  {NoStop}%
\bibitem [{\citenamefont {Yu}\ and\ \citenamefont {Cardona}(1995)}]{yu1995}%
  \BibitemOpen
  \bibfield  {author} {\bibinfo {author} {\bibfnamefont {P.}~\bibnamefont
  {Yu}}\ and\ \bibinfo {author} {\bibfnamefont {M.}~\bibnamefont {Cardona}},\
  }\href@noop {} {\emph {\bibinfo {title} {Fundamentals of Semiconductors}}}\
  (\bibinfo  {publisher} {Springer},\ \bibinfo {address} {Heidelberg},\
  \bibinfo {year} {1995})\BibitemShut {NoStop}%
\bibitem [{\citenamefont {Friedland}\ \emph {et~al.}(1996)\citenamefont
  {Friedland}, \citenamefont {Hey}, \citenamefont {Kostial}, \citenamefont
  {Klann},\ and\ \citenamefont {Ploog}}]{friedland1996}%
  \BibitemOpen
  \bibfield  {author} {\bibinfo {author} {\bibfnamefont {K.~J.}\ \bibnamefont
  {Friedland}}, \bibinfo {author} {\bibfnamefont {R.}~\bibnamefont {Hey}},
  \bibinfo {author} {\bibfnamefont {H.}~\bibnamefont {Kostial}}, \bibinfo
  {author} {\bibfnamefont {R.}~\bibnamefont {Klann}}, \ and\ \bibinfo {author}
  {\bibfnamefont {K.}~\bibnamefont {Ploog}},\ }\href@noop {} {\bibfield
  {journal} {\bibinfo  {journal} {Phys. Rev. Lett.}\ }\textbf {\bibinfo
  {volume} {77}},\ \bibinfo {pages} {4616} (\bibinfo {year}
  {1996})}\BibitemShut {NoStop}%
\bibitem [{\citenamefont {Friedland}\ \emph {et~al.}(2001)\citenamefont
  {Friedland}, \citenamefont {Riedel}, \citenamefont {Kostial}, \citenamefont
  {H\"oricke}, \citenamefont {Hey},\ and\ \citenamefont
  {Ploog}}]{friedland2001}%
  \BibitemOpen
  \bibfield  {author} {\bibinfo {author} {\bibfnamefont {K.~J.}\ \bibnamefont
  {Friedland}}, \bibinfo {author} {\bibfnamefont {A.}~\bibnamefont {Riedel}},
  \bibinfo {author} {\bibfnamefont {H.}~\bibnamefont {Kostial}}, \bibinfo
  {author} {\bibfnamefont {M.}~\bibnamefont {H\"oricke}}, \bibinfo {author}
  {\bibfnamefont {R.}~\bibnamefont {Hey}}, \ and\ \bibinfo {author}
  {\bibfnamefont {K.~H.}\ \bibnamefont {Ploog}},\ }\href@noop {} {\bibfield
  {journal} {\bibinfo  {journal} {J. Electron. Mater.}\ }\textbf {\bibinfo
  {volume} {90}},\ \bibinfo {pages} {817} (\bibinfo {year} {2001})}\BibitemShut
  {NoStop}%
\bibitem [{\citenamefont {Nikulin}(1998)}]{nikulin1998}%
  \BibitemOpen
  \bibfield  {author} {\bibinfo {author} {\bibfnamefont {A.~Y.}\ \bibnamefont
  {Nikulin}},\ }\href@noop {} {\bibfield  {journal} {\bibinfo  {journal} {Phys.
  Rev. B}\ }\textbf {\bibinfo {volume} {57}},\ \bibinfo {pages} {11178}
  (\bibinfo {year} {1998})}\BibitemShut {NoStop}%
\bibitem [{\citenamefont {Nikulin}\ and\ \citenamefont
  {Zaumseil}(1999)}]{nikulin1999}%
  \BibitemOpen
  \bibfield  {author} {\bibinfo {author} {\bibfnamefont {A.~Y.}\ \bibnamefont
  {Nikulin}}\ and\ \bibinfo {author} {\bibfnamefont {P.}~\bibnamefont
  {Zaumseil}},\ }\href@noop {} {\bibfield  {journal} {\bibinfo  {journal}
  {Phys. Status Solidi A}\ }\textbf {\bibinfo {volume} {172}},\ \bibinfo
  {pages} {291} (\bibinfo {year} {1999})}\BibitemShut {NoStop}%
\bibitem [{\citenamefont {Speriosu}\ and\ \citenamefont
  {Vreeland}(1984)}]{speriosu1984}%
  \BibitemOpen
  \bibfield  {author} {\bibinfo {author} {\bibfnamefont {V.~S.}\ \bibnamefont
  {Speriosu}}\ and\ \bibinfo {author} {\bibfnamefont {T.}~\bibnamefont
  {Vreeland}},\ }\href@noop {} {\bibfield  {journal} {\bibinfo  {journal} {J.
  Appl. Phys.}\ }\textbf {\bibinfo {volume} {56}},\ \bibinfo {pages} {1591}
  (\bibinfo {year} {1984})}\BibitemShut {NoStop}%
\bibitem [{\citenamefont {Holy}\ \emph {et~al.}(1992)\citenamefont {Holy},
  \citenamefont {Kubena}, \citenamefont {Ohlidal},\ and\ \citenamefont
  {Ploog}}]{holy1992}%
  \BibitemOpen
  \bibfield  {author} {\bibinfo {author} {\bibfnamefont {V.}~\bibnamefont
  {Holy}}, \bibinfo {author} {\bibfnamefont {J.}~\bibnamefont {Kubena}},
  \bibinfo {author} {\bibfnamefont {I.}~\bibnamefont {Ohlidal}}, \ and\
  \bibinfo {author} {\bibfnamefont {K.~H.}\ \bibnamefont {Ploog}},\ }\href@noop
  {} {\bibfield  {journal} {\bibinfo  {journal} {Superlatt. Microstr.}\
  }\textbf {\bibinfo {volume} {12}},\ \bibinfo {pages} {25} (\bibinfo {year}
  {1992})}\BibitemShut {NoStop}%
\bibitem [{\citenamefont {Proctor}\ \emph {et~al.}(1994)\citenamefont
  {Proctor}, \citenamefont {Oelgart}, \citenamefont {Rhan},\ and\ \citenamefont
  {Reinhart}}]{proctor1994}%
  \BibitemOpen
  \bibfield  {author} {\bibinfo {author} {\bibfnamefont {M.}~\bibnamefont
  {Proctor}}, \bibinfo {author} {\bibfnamefont {G.}~\bibnamefont {Oelgart}},
  \bibinfo {author} {\bibfnamefont {H.}~\bibnamefont {Rhan}}, \ and\ \bibinfo
  {author} {\bibfnamefont {F.~K.}\ \bibnamefont {Reinhart}},\ }\href@noop {}
  {\bibfield  {journal} {\bibinfo  {journal} {Appl. Phys. Lett.}\ }\textbf
  {\bibinfo {volume} {64}},\ \bibinfo {pages} {3154} (\bibinfo {year}
  {1994})}\BibitemShut {NoStop}%
\bibitem [{\citenamefont {Pan}\ \emph {et~al.}(1999)\citenamefont {Pan},
  \citenamefont {Wang}, \citenamefont {Zhuang}, \citenamefont {Lin},
  \citenamefont {Zhou}, \citenamefont {Li}, \citenamefont {Wu},\ and\
  \citenamefont {Wang}}]{pan1999}%
  \BibitemOpen
  \bibfield  {author} {\bibinfo {author} {\bibfnamefont {Z.}~\bibnamefont
  {Pan}}, \bibinfo {author} {\bibfnamefont {Y.~T.}\ \bibnamefont {Wang}},
  \bibinfo {author} {\bibfnamefont {Y.}~\bibnamefont {Zhuang}}, \bibinfo
  {author} {\bibfnamefont {Y.~W.}\ \bibnamefont {Lin}}, \bibinfo {author}
  {\bibfnamefont {Z.~Q.}\ \bibnamefont {Zhou}}, \bibinfo {author}
  {\bibfnamefont {L.~H.}\ \bibnamefont {Li}}, \bibinfo {author} {\bibfnamefont
  {R.~H.}\ \bibnamefont {Wu}}, \ and\ \bibinfo {author} {\bibfnamefont {Q.~M.}\
  \bibnamefont {Wang}},\ }\href@noop {} {\bibfield  {journal} {\bibinfo
  {journal} {Appl. Phys. Lett.}\ }\textbf {\bibinfo {volume} {75}},\ \bibinfo
  {pages} {223} (\bibinfo {year} {1999})}\BibitemShut {NoStop}%
\bibitem [{\citenamefont {Zachariasen}(1994)}]{zachariasen1945}%
  \BibitemOpen
  \bibfield  {author} {\bibinfo {author} {\bibfnamefont {W.~H.}\ \bibnamefont
  {Zachariasen}},\ }\href@noop {} {\emph {\bibinfo {title} {Theory of X-ray
  Diffraction in Crystals}}}\ (\bibinfo  {publisher} {Dover Publications
  Inc.},\ \bibinfo {address} {New York},\ \bibinfo {year} {1994})\BibitemShut
  {NoStop}%
\bibitem [{\citenamefont {Elmer}(2012)}]{elmer2012}%
  \BibitemOpen
  \bibfield  {author} {\bibinfo {author} {\bibnamefont {Elmer}},\ }\href@noop
  {} {\emph {\bibinfo {title} {Open Source Finite Element Software for
  Multiphysical Problems}}}\ (\bibinfo  {publisher}
  {http://www.csc.fi/english/pages/elmer},\ \bibinfo {address} {Espoo},\
  \bibinfo {year} {2012})\BibitemShut {NoStop}%
\bibitem [{\citenamefont {Hey}\ \emph {et~al.}(2009)\citenamefont {Hey},
  \citenamefont {Ramsteiner}, \citenamefont {Santos},\ and\ \citenamefont
  {Friedland}}]{Hey2009}%
  \BibitemOpen
  \bibfield  {author} {\bibinfo {author} {\bibfnamefont {R.}~\bibnamefont
  {Hey}}, \bibinfo {author} {\bibfnamefont {M.}~\bibnamefont {Ramsteiner}},
  \bibinfo {author} {\bibfnamefont {P.}~\bibnamefont {Santos}}, \ and\ \bibinfo
  {author} {\bibfnamefont {K.~J.}\ \bibnamefont {Friedland}},\ }\href@noop {}
  {\bibfield  {journal} {\bibinfo  {journal} {J. Cryst. Growth}\ }\textbf
  {\bibinfo {volume} {311}},\ \bibinfo {pages} {1680} (\bibinfo {year}
  {2009})}\BibitemShut {NoStop}%
\bibitem [{\citenamefont {Coleridge}(1991)}]{Coleridge1991}%
  \BibitemOpen
  \bibfield  {author} {\bibinfo {author} {\bibfnamefont {P.~T.}\ \bibnamefont
  {Coleridge}},\ }\href@noop {} {\bibfield  {journal} {\bibinfo  {journal}
  {Phys. Rev. B}\ }\textbf {\bibinfo {volume} {44}},\ \bibinfo {pages} {3793}
  (\bibinfo {year} {1991})}\BibitemShut {NoStop}%
\bibitem [{\citenamefont {Mancoff}\ \emph {et~al.}(1996)\citenamefont
  {Mancoff}, \citenamefont {Zielinski}, \citenamefont {Marcus}, \citenamefont
  {Campman},\ and\ \citenamefont {Gossard}}]{Mancoff1996}%
  \BibitemOpen
  \bibfield  {author} {\bibinfo {author} {\bibfnamefont {F.~B.}\ \bibnamefont
  {Mancoff}}, \bibinfo {author} {\bibfnamefont {L.~J.}\ \bibnamefont
  {Zielinski}}, \bibinfo {author} {\bibfnamefont {C.~M.}\ \bibnamefont
  {Marcus}}, \bibinfo {author} {\bibfnamefont {K.}~\bibnamefont {Campman}}, \
  and\ \bibinfo {author} {\bibfnamefont {A.~C.}\ \bibnamefont {Gossard}},\
  }\href@noop {} {\bibfield  {journal} {\bibinfo  {journal} {Phys. Rev. B}\
  }\textbf {\bibinfo {volume} {53}},\ \bibinfo {pages} {7599} (\bibinfo {year}
  {1996})}\BibitemShut {NoStop}%
\bibitem [{\citenamefont {Chaplik}\ and\ \citenamefont
  {Chaplik}(2000)}]{Chaplik2000}%
  \BibitemOpen
  \bibfield  {author} {\bibinfo {author} {\bibfnamefont {A.}~\bibnamefont
  {Chaplik}}\ and\ \bibinfo {author} {\bibfnamefont {V.}~\bibnamefont
  {Chaplik}},\ }\href@noop {} {\bibfield  {journal} {\bibinfo  {journal} {JETP
  Lett.}\ }\textbf {\bibinfo {volume} {72}},\ \bibinfo {pages} {503} (\bibinfo
  {year} {2000})}\BibitemShut {NoStop}%
\bibitem [{\citenamefont {Friedland}\ \emph {et~al.}(2010)\citenamefont
  {Friedland}, \citenamefont {Hey}, \citenamefont {Riedel},\ and\ \citenamefont
  {Maude}}]{Friedland2010}%
  \BibitemOpen
  \bibfield  {author} {\bibinfo {author} {\bibfnamefont {K.~J.}\ \bibnamefont
  {Friedland}}, \bibinfo {author} {\bibfnamefont {R.}~\bibnamefont {Hey}},
  \bibinfo {author} {\bibfnamefont {A.}~\bibnamefont {Riedel}}, \ and\ \bibinfo
  {author} {\bibfnamefont {D.~K.}\ \bibnamefont {Maude}},\ }\href@noop {}
  {\bibfield  {journal} {\bibinfo  {journal} {Phys. Status Solidi C}\ }\textbf
  {\bibinfo {volume} {7}},\ \bibinfo {pages} {2562} (\bibinfo {year}
  {2010})}\BibitemShut {NoStop}%
\bibitem [{\citenamefont {Harrang}\ \emph {et~al.}(1985)\citenamefont
  {Harrang}, \citenamefont {Higgins}, \citenamefont {Goodall}, \citenamefont
  {Jay}, \citenamefont {Laviron},\ and\ \citenamefont
  {Delescluse}}]{Harrang1985}%
  \BibitemOpen
  \bibfield  {author} {\bibinfo {author} {\bibfnamefont {J.~P.}\ \bibnamefont
  {Harrang}}, \bibinfo {author} {\bibfnamefont {R.~J.}\ \bibnamefont
  {Higgins}}, \bibinfo {author} {\bibfnamefont {R.~K.}\ \bibnamefont
  {Goodall}}, \bibinfo {author} {\bibfnamefont {P.~R.}\ \bibnamefont {Jay}},
  \bibinfo {author} {\bibfnamefont {M.}~\bibnamefont {Laviron}}, \ and\
  \bibinfo {author} {\bibfnamefont {P.}~\bibnamefont {Delescluse}},\
  }\href@noop {} {\bibfield  {journal} {\bibinfo  {journal} {Phys. Rev. B}\
  }\textbf {\bibinfo {volume} {32}},\ \bibinfo {pages} {8126} (\bibinfo {year}
  {1985})}\BibitemShut {NoStop}%
\end{thebibliography}
%


\begin{table*}[htbp]
  \caption{Thicknesses, in-plane epitaxial strain ($\epsilon_{\text{epi} }$), and elastic contants  used in the strain calculations for Fig.~\ref{calc}. The elastic constants of the alloy layers were obtained by interpolating the values for the constituent materials GaAs, AlAs, and InAs .}

    \begin{tabular}{l c c c c}
    \toprule
          &    thickness    &    epi-strain    & $   c_{11}   $    &    $   c_{12}   $ \\

          & (nm)  & $\epsilon_{\text{epi} }$ & (N/m$^2$) & (N/m$^2$) \\
    \hline
    \multicolumn{1}{c}{In$_{0.24}$Ga$_{0.76}$As} & 15    & ~~$-$1.68$\times$10$^{-2}$~~ &~~ 1.10$\times$10$^{11}$~~ & ~~5.13$\times$10$^{10}$~~ \\
    \multicolumn{1}{c}{Al$_{0.33}$Ga$_{0.67}$As} & 62    & $-$4.58$\times$10$^{-4}$ & 1.19$\times$10$^{11}$ & 5.45$\times$10$^{10}$ \\
    \multicolumn{1}{c}{GaAs QW} & 13    & 0 & 1.18$\times$10$^{11}$ & 5.32$\times$10$^{10}$ \\
    \multicolumn{1}{c}{Al$_{0.33}$Ga$_{0.67}$As} & 62    & $-$4.58$\times$10$^{-4}$ & 1.19$\times$10$^{11}$ & 5.45$\times$10$^{10}$ \\
        \toprule
    \end{tabular}%
  \label{tab:addlabel}%
\end{table*}%
\begin{table*}[ht]
\caption{Cancelation of SdH oscillations}

\begin{tabular}{l c c c c c} 
\hline\hline 
Hall bar                & density            & mobility   & $\tau_{\text{tr}}$     & $\tau_{\text{tot}}$    & $B_{\text{onset}}$ \\ 
                        & $(10^{15}$~m$^{-2})$  & (m$^2V^{-1}s^{-1}$) & $(10^{-12}$~sec) & $(10^{-12}$~sec) & (T)     \\ 
\hline 
flat                    & 6.32               & 126        & 50              & 1.15            & 0.19        \\ 
flat-average $20~\mu$m & 6.32               & 126        & 50              & $\sim 0.9$      & 0.19        \\
\hline
curved $10~\mu$m        & 6.28               & 118        & 46              & 0.43            & 0.56        \\
curved $20~\mu$m        & 6.28               & 118        & 46              & 0.43            & 0.57        \\ 
\hline 
\label{table1} 
\end{tabular}

\end{table*}

\end{document}